\documentclass[aps, prd, twocolumn, preprintnumbers, superscriptaddress, nofootinbib, floatfix]{revtex4-1}

\usepackage{amsmath, amsfonts}
\usepackage{bm}
\usepackage{graphicx}
\usepackage{epstopdf}
\usepackage{hyperref}
\usepackage{array}
\usepackage{soul}
\usepackage{color}

\newcommand{\exclude}[1]{}

\enlargethispage{\baselineskip}

\hypersetup{
     colorlinks   = true,
     citecolor    = blue
}

\everymath{\displaystyle}

\newcommand{\Mp}{m_{\rm Pl}}
\newcommand{\MS}{M_{\odot}}

\newcommand{\ns}{n_A^{\rm mis}}

\newcommand{\rmt}{\rho_A^{\rm mis}(t_0)}
\newcommand{\rst}{\rho_A^{\rm str}(t_0)}
\newcommand{\rwt}{\rho_A^{\rm wall}(t_0)}
\newcommand{\as}{\alpha_{\rm str}}
\newcommand{\aw}{\alpha_{\rm wall}}
\newcommand{\at}{\alpha_{\rm tot}}

\newcommand{\mCDM}{m_{\rm CDM}}
\newcommand{\MW}{M_{\rm MW}}
\newcommand{\TM}{T_{\rm MeV}}
\newcommand{\tS}{\tau_{\odot}}
\newcommand{\nS}{n_c}

\newcommand{\Ms}{M_{\odot}}
\newcommand{\rhS}{\rho_{\odot}}
\newcommand{\rS}{r_{\odot}}
\newcommand{\vS}{v_{\odot}}
\newcommand{\HRH}{H_{\rm RH}}

\newcommand{\Teq}{T_{\rm eq}}

\newcommand{\TL}{T_\Lambda}

\newcommand{\TRH}{T_{\rm RH}}

\newcommand{\aeq}{a_{\rm eq}}
\newcommand{\aRH}{a_{\rm RH}}

\newcommand{\tRH}{t_{\rm RH}}

\newcommand{\bx}{{\bf x}}

\newcommand{\std}{{\rm std}}
\newcommand{\RCDM}{\rho_{\rm CDM}}

\renewcommand\({\left(}
\renewcommand\){\right)}

\newcommand{\be}{\begin{equation}}
\newcommand{\ee}{\end{equation}}
\newcommand{\bea}{\begin{eqnarray}}
\newcommand{\eea}{\end{eqnarray}}

\begin{document}

\title{\bf Axion Miniclusters in Modified Cosmological Histories}

 \newcommand{\FIRSTAFF}{\affiliation{Department of Physics and Astronomy,
	Uppsala University,
	L\"agerhyddsv\"agen 1,
	75120 Uppsala,
	Sweden}}
\newcommand{\SECONDAFF}{\affiliation{Nordita,
	KTH Royal Institute of Technology and Stockholm University,
	Roslagstullsbacken 23,
	10691 Stockholm,
	Sweden}}
\newcommand{\THIRDAFF}{\affiliation{Gravitation Astroparticle Physics Amsterdam (GRAPPA), \\ 
	Institute for Theoretical Physics Amsterdam and Delta Institute for Theoretical Physics,\\ 
	University of Amsterdam, Science Park 904, 1098 XH Amsterdam, The Netherlands}	}
\newcommand{\FOURTHAFF}{\affiliation{Departamento de F\'isica Te\'orica, 
			Universidad de Zaragoza. 
			c/ Pedro Cerbuna 12,
			50009 Zaragoza,
			Spain}}
\newcommand{\FIFTHAFF}{\affiliation{Max Planck Institute for Physics,
			Foehringer Ring 6,
			80805 Munich,
			Germany}}

\author{Luca Visinelli}
\email[Electronic address: ]{l.visinelli@uva.nl}
\FIRSTAFF
\SECONDAFF
\THIRDAFF
\author{Javier Redondo}
\email[Electronic address: ]{jredondo@unizar.es}
\FOURTHAFF
\FIFTHAFF

\date{\today}
\preprint{NORDITA-2018-063, MPP-2018-237}

\begin{abstract}
If the symmetry breaking leading to the origin of the axion dark matter field occurs after the end of inflation and is never restored, then overdensities in the axion field collapse to form dense objects known in the literature as axion miniclusters. The estimates of the typical minicluster mass and radius strongly depend on the details of the cosmology at which the onset of axion oscillations begin. In this work we study the properties and phenomenology of miniclusters in alternative cosmological histories and find that they can change by many orders of magnitude. Our findings have direct implications on current and future experimental searches and, in the case of discovery, could be used to learn something about the universe expansion prior to Big-Bang-Nucleosynthesis.
\end{abstract}

\maketitle

\section{Introduction}

The nature of the cold dark matter (CDM) remains unknown to date despite the growth of evidence in support of its existence coming, on top of the original motivations~\cite{1976AJ.....81..687R, 1976AJ.....81..719R}, from gravitational lensing~\cite{Trimble:1987}, the cosmic microwave background radiation (CMBR)~\cite{Ade:2015xua, Aghanim:2018eyx}, also in combination with Lyman-$\alpha$ and weak lensing~\cite{Lesgourgues:2007te}, the hierarchical structure formation of the observable universe~\cite{Springel:2005nw}, the formation and evolution of galaxies~\cite{Davis:1985rj, Efstathiou:1985re, Springel:2006vs}, galactic collisions~\cite{Clowe:2003tk, Markevitch:2003at}, and a plethora of other observational techniques.

Among the many hypothetical particles that could compose the CDM is the quantum chromodynamics (QCD) axion~\cite{Weinberg:1977ma, Wilczek:1977pj}. The axion is the pseudo-Goldstone boson arising in the spontaneous breaking of a U(1) symmetry first introduced by Peccei and Quinn (PQ~\cite{Peccei:1977hh, Peccei:1977ur}) to address the strong-CP problem~\cite{Belavin:1975fg, tHooft:1976rip, Jackiw:1976pf, Callan:1976je}. The fact that axions could solve two distinct problems in physics makes its search particularly appealing. If the axion field exists, it has a very small mass and faint couplings to ordinary particles. Both happen to be suppressed by a new energy scale, the axion decay constant $f_A$, which corresponds approximately to the scale of PQ symmetry breaking and which is constrained by axion phenomenology to be $f_A\gtrsim 10^7\,$GeV. In particular, the scale $f_A^{-1}$ sets the axion coupling to two photons, which opens the possibility for axion electrodynamics~\cite{Wilczek:1987mv, Krasnikov:1996bm, Li:2009tca, visinelli:2013fia, Tercas:2018gxv, Visinelli:2018zif} and promising laboratory detection methods~\cite{Stern:2016bbw, Raggi:2014zpa, Majorovits:2016yvk, Kahn:2016aff, Alesini:2017ifp, Alesini:2019nzq}. In the literature, mixed dark matter models in which the axion makes up a fraction of the dark matter while the rest is in the form of weakly interacting massive particles (WIMPs) have also been considered~\cite{Bae:2014efa, Bae:2015rra, Baum:2016oow}. See Refs.~\cite{Raffelt:1995ym, Raffelt:2006rj, Sikivie:2006ni, Kim:2008hd, Wantz:2009it, Kawasaki:2013ae, Marsh:2015xka,Kim:2017yqo} for reviews of the QCD axion.

The history and the properties of the present axion field strongly depend on the moment at which the breaking of the PQ symmetry occurs with respect to inflation~\cite{Linde:1987bx, Linde:1991km, Turner:1991, Wilczek:2004cr, Tegmark:2005dy, Hertzberg:2008wr, Freivogel:2008qc, Mack:2009hv, Visinelli:2009zm, Acharya:2010zx, Ballesteros:2016xej, Hoof:2018ieb, Tenkanen:2019xzn}. If the PQ symmetry breaking occurs after inflation, a fraction of the total axions component is expected to organize into gravitationally bound structures known as axion ``miniclusters''~\cite{Hogan:1988mp, Kolb:1993zz, Kolb:1993hw, Kolb:1994fi, Sakharov:1996xg}, prompted by the inhomogeneities of the axion field in this scenario. Axion miniclusters are compact objects with a density of various orders of magnitude higher than the present local CDM density. Inside axion miniclusters another type of exotic structure, an axion star~\cite{Kaup:1968zz, Ruffini:1969qy, Das_1963, Feinblum:1968nwc, Teixeira:1975ad, Colpi:1986ye, Seidel:1991zh, Tkachev:1991ka, Chavanis:2011, Braaten:2015eeu, Eby:2015hyx, Eby:2016cnq, Levkov:2016rkk, Helfer:2016ljl, Braaten:2016dlp, Braaten:2016kzc, Bai:2016wpg, Eby:2017xaw, Desjacques:2017fmf, Visinelli:2017ooc, Chavanis:2017loo, Krippendorf:2018tei, Eggemeier:2019jsu}, could possibly form. It has been argued that the first miniclusters that ever come into place have a characteristic size of the order of $\sim 10^{-12}$ solar masses. This scale is much smaller than the smallest clump that WIMPs would assemble into in the standard cosmology, because of their much longer free-streaming length~\footnote{See Ref.~\cite{Gelmini:2008sh} for the effect of a non-standard cosmology on the free-streaming length of WIMPs.}. Therefore, detecting these clumps provides a unique discrimination signature among CDM candidates. As structure formation evolves, axion miniclusters are expected to hierarchically assemble into dark matter halos of galactic size, forming minicluster halos. This claim has yet to be addressed numerically, as well as the possibility that miniclusters might have not survived tidal disruption. Some studies suggest that miniclusters survive hierarchical structure formation to date~\cite{Zurek:2006sy, Tinyakov:2015cgg, Hardy:2016mns, Dokuchaev2017}, claiming that it is possible to constrain the fraction of dark matter in halos using micro-lensing data~\cite{Fairbairn:2017dmf, Fairbairn:2017sil} and femto-lensing in the future~\cite{Katz:2018zrn} (see also \cite{Inomata:2017vxo}). Semi-analytic results on the mass function of axion miniclusters are available today~\cite{Enander:2017ogx}, with refined numerical work recently reported~\cite{Vaquero:2018tib, Eggemeier:2019khm}.

Today, the standard axion miniclusters would be gravitationally bound clumps of axions with masses of the order of the largest asteroids like Vesta or Pallas, and of size comparable to an astronomical unit. Miniclusters could suffer tidal disruption by stars, with the value of bound axions diminishing with respect to its early universe estimate. It is usually expected that a sizeable fraction of the galactic DM axions is bound into minicluster structures, the remaining part forming a relatively homogeneous halo. Despite the large number of clumps expected, the Earth would encounter only a few such objects every galactic year (!). On the other hand, if the fraction of axions bound into miniclusters is close to unity, the direct detection of axions in microwave cavity searches could be severely affected.  A negative search by a cavity experiment could be an indication that axions are mostly organised into clumped structures like miniclusters. 
Since axion detection is sensitive to the local CDM energy density, a clumpy axion distribution would lead to spikes in the axion detection spectrum and be relevant for a direct detection technique. The interest in all of the upcoming axion detectors then lies in the present phase-space distribution of the axion CDM, which is not expected to be homogeneous even at the interstellar scale.
A reliable detection must take into account the possibility of a inhomogeneous CDM distribution either in space (axion miniclusters and stars) or in momentum (low-dispersion filaments from tidal stripping). 
Moreover, if the axion is discovered by a haloscope experiment like ADMX, HAYSTAC or CULTASK, the energy spectrum will be immediately measured and could be used to do galactic astronomy~\cite{OHare:2017yze}. The spectrum and its variation in time (daily and anual modulations) can be used to identify substructure in the axion DM distribution like tidal streams from dwarf-galaxies~\cite{Vogelsberger:2010gd} or even from axion miniclusters themselves, obtaining their main properties with precision~\cite{OHare:2017yze}. With these techniques, astronomical quantities like the solar peculiar velocity could be measured even better than with ordinary astronomy. Several variations of the haloscope concept allow to measure the axion velocity distribution by enabling directional detection~\cite{Irastorza:2012jq, Irastorza:2018dyq, Knirck:2018knd}, increasing the precision and decreasing the required measurement time. Understanding the fraction of axions bound in miniclusters and of the axion phase-space will maximise the outcome of the various experiments that will start looking for axion CDM in the near future.

The rough properties of axion miniclusters described above have been studied only under the assumption that the universe is radiation-dominated when axions become non-relativistic. This is certainly a simple and minimal assumption but also one that does not need to be necessarily correct. We know with some certainty that the universe expansion must be radiation-dominated after and around neutrino decoupling (temperatures of $T\lesssim 5\,$MeV) not to alter the successful predictions of radiation-dominated Big-Bang Nucleosynthesis (BBN)~\cite{Kawasaki:1999na, Kawasaki:2000en, Hannestad:2004px, Ichikawa:2005vw, DeBernardis:2008zz}, but we have no direct evidence of the expansion rate of the universe at earlier times (higher temperatures). Most importantly, if axions are to be a dominant contribution to the CDM they become CDM at temperatures $\sim\,$GeV, precisely when we cannot ascertain the crucial assumption of radiation domination. 
 
In this paper we want to drop entirely this assumption and study the properties of axion miniclusters in different non-standard cosmologies (NSCs) before BBN. The mass and the radius of a minicluster depend crucially on the size of the causal horizon at the time when axions become non-relativistic, which in alternative cosmologies might differ by various orders of magnitude with respect to the standard scenario. 
In order to focus the discussion on the novel aspects, we assume that axions make up the totality of the cold dark matter observed and that essentially all axions fall into miniclusters. Our results, summarised in \mbox{Table~\ref{table_parameters}} are quite spectacular! In fact, the typical minicluster mass and radius can change by many orders of magnitude with respect to the predictions achieved in the standard cosmology. Most importantly, the time and duration of encounters with the Earth can be largely enhanced or suppressed, opening many possibilities for the direct detection of axion dark matter.

Previous works have already studied how the axion mass for which the axion explains the totality of the observed CDM is modified in several NSCs~\cite{Visinelli:2009kt, Visinelli:2017imh, Ramberg:2019dgi} but the changes to the properties of miniclusters are presented here for the first time. Moreover, regarding the treatment of NSCs and the axion DM mass, we improve over previous results by computing the non-standard cosmologies numerically including the recent detailed input from lattice QCD (equation of state and axion mass)~\cite{Borsanyi:2016ksw} and presenting simple analytical comparisons with the standard radiation dominated case. Note that, in addition to the cold axion population, a modified cosmology would alter the yield of thermally produced axions~\cite{Grin:2007yg}, as well as of any other light particle such as neutrinos~\cite{Giudice:2000dp, Giudice:2001ep}; however, since for the range of masses considered in this work, $m_A \lesssim 10\,$meV, the thermal population is a sub-dominant component to the total energy density, we do not discuss this contribution further.

This paper is organised as follows. Sec.~\ref{axion_review} is a brief review of the production of cosmological axions. In Sec.~\ref{cosmological_modified} we set the stage for the non-standard cosmological models of interest and we review their impact on the DM axion mass. Sec.~\ref{miniclusters_analytic} presents our results on the properties of axion miniclusters and their phenomenology. Final remarks and conclusions are drawn in Sec.~\ref{Discussion and conclusions}.

\section{Axion cold dark matter} \label{axion_review}

In the post-inflationary scenario, axion cold dark matter is produced from the misalignment mechanism and the decay of topological defects (strings and walls). The axion field behaves like a massless field, becoming homogenous on scales of the order of the causal horizon, as long as the Hubble expansion rate ($H=d\log a/dt$, with $a=a(t)$ the scale factor of a Friedmann-Robertson-Walker universe) is much larger than the axion mass, $H \gg m_A$. However, around the time $t_1$ (or equivalently the temperature $T_1$) when the condition
\be
	3 H(T_1) \approx m_A(T_1) , 
	\label{eq:111}
\ee
is met, the relevant long-wavelength modes begin to respond to the QCD potential by evolving towards the minimum of the field configuration $A = 0$ (where $A(x)$ is the axion field), oscillating around it as a non-relativistic (NR) field with equation of state $w_A\to 0$, {\it i.e.} equivalent to CDM. Shortly after that time, the energy density distribution, which corresponds to the axion field oscillating with different amplitudes at different positions, becomes essentially frozen. The calculation of the spatially-averaged CDM yield from the misalignment mechanism is reviewed in Appendix~\ref{Production mechanisms} and it is given by
\begin{eqnarray} \nonumber
	\rho_A^{\rm mis} &=& \frac{1}{2} m_A(T_0)m_A(T_1) f_A^2 \langle \theta_i^2 \rangle \(\frac{a_1}{a_0}\)^3 = \\ 
	&=& \frac{1}{2} \sqrt{\chi(T_0)\chi(T_1)} \langle \theta_i^2 \rangle \(\frac{a_1}{a_0}\)^3.
	\label{energy_density_t1}
\end{eqnarray}
In the last expression, we have introduced the temperature-dependent axion mass $m_A(T)$, which is related to the axion decay constant $f_A$ by
\be
	m_A^2(T)\,f_A^2 = \chi(T),
\ee
where $\chi(T)$ is the topological susceptibility of QCD. For practical purposes it is useful to define the energy scale $\Lambda_A = (\chi(T=0))^{1/4} \approx 75.5\,$MeV in terms of the susceptibility at zero temperature~\cite{diCortona:2015ldu,Borsanyi:2016ksw}. In Eq.~\eqref{energy_density_t1} we introduced the standard notation, in which we have normalised the axion field with $f_A$ by defining the angle $\theta = A/f_A \in(-\pi,\pi)$, and we wrote $\langle\theta_i^2\rangle$ for the average of the initial misalignment angle squared, including the contribution from the non-harmonic terms in the axion potential in Eq.~\eqref{Vqcd}~\cite{Bae:2008ue, Visinelli:2009zm, Wantz:2009it, Visinelli:2014twa}. Finally, the last factor in Eq.~\eqref{energy_density_t1} involving the ratio of the scale factor at $t_1$ and today $t_0$, is the dilution of the axion number density due to the expansion of the universe from the moment at which the axions become NR and behave as CDM until present time. Note that the energy density in Eq.~\eqref{energy_density_t1} is a quantity uniquely related to QCD scales: the axion decay constant enters the computation through the determination of the temperature $T_1$ at the onset of the field oscillations in Eq.~\eqref{eq:111}, or equivalently the scale factor $a_1 = a(T_1)$. The quantity $T_1$ is sensitive to the energy content of the universe, therefore any deviation from the standard cosmological model affects the axion DM yield and the size of the horizon ($\sim 1/H(t_1)\equiv 1/H_1$) when axions start to behave as CDM.

DM axions are also produced from the decay of strings and domain walls, with energy density contributions $\rho_A^{\rm str}$ and $\rho_A^{\rm wall}$ respectively that scale with the same exponent in the decay constant $f_A$~\cite{Sikivie:2006ni, Hiramatsu:2010yn, Klaer:2017ond}, at least when the details of the production and the scaling with the power spectra are neglected. This allows us to parametrise the contribution from the topological defects to the axion cold dark matter density as an additional factor multiplying the misalignment contribution,
\begin{equation}
	\bar\rho_A = \rho_A^{\rm wall} + \rho_A^{\rm str} + \rho^{\rm mis} \equiv \at\rho_A^{\rm mis}.
	\label{eq:totalaxionenergydensity}
\end{equation}
This parametrisation allows for the case in which the axion CDM yield is even smaller than the misalignment-only contribution $\at < 1$, a scenario supported by recent simulations~\cite{Fleury:2015aca, Klaer:2017qhr, Gorghetto:2018myk}.

\section{Axion population in modified cosmological scenarios} \label{cosmological_modified}

\subsection{Parametrising a cosmological model} \label{generic_model}

In this paper we consider a non-standard cosmology (NSC), a scenario in which the early universe, close to the relevant temperature $T_1\lesssim \mathcal{O}(1)\,$GeV, contains the standard thermal bath of SM particles and an additional component labelled $\phi$, whose energy density is so large during the temperatures considered that it affects the expansion rate of the universe. The evolution of the energy density of radiation ($\rho_R$) and of the extra substance ($\rho_\phi$) is governed by a set of coupled Boltzmann equations,
\bea
	\dot \rho_\phi + 3(1+w_\phi)H\rho_\phi &=& -\Gamma_\phi \rho_\phi , \label{eq:boltzmannpsi}\\
	\dot \rho_R + 3(1+w_R)H\rho_R &=& \Gamma_\phi \rho_\phi,\label{eq:boltzmannR}
\eea
where a dot denotes a derivative with respect to cosmic time $t$, the parameter $w_i = p_i/\rho_i$ describes the equation of state of substance $i$ with pressure $p_i$ and energy density $\rho_i$ (with $i \in \{\phi, R\}$), and $\Gamma_\phi$ is a possible decay rate. The Friedmann equation $H^2= (8\pi/3\Mp^2)(\rho_R+\rho_\phi)$ gives the evolution of the scale factor (and thus $H$) as a function of time. In order to connect with a late cosmology in agreement with observations we will consider substances that decay and/or redshift faster than radiation, so that the description of the history of the universe transitions from the NSC to the standard scenario after a certain reheat time ($\tRH$) or equivalently below a certain temperature, which we indicate as the reheating temperature $T_{\rm RH}$. We define the reheat time implicitly by the condition $\rho_\phi(\tRH) = \rho_R(\tRH)$. The Hubble rate at time $\tRH$ is then
\be
	\HRH \!=\! \sqrt{\frac{8\pi}{3}\frac{\rho_\phi(\tRH) \!+\! \rho_R(\tRH)}{\Mp^2}} \!\approx\! \sqrt{\frac{8\pi^3}{45}\,g_*(\TRH)}\,\frac{\TRH^2}{\Mp}.
	\label{eq:Hubble_reheat}
\ee
In the last step we have assumed that the extra energy injected reaches the thermal equilibrium on a much smaller time-scale than the Hubble expansion rate, so that the contribution from the radiation to the energy density is $\rho_R = (\pi^2/30)g_*(T)T^4$.

Once the extra substance has decayed, its energy density $\rho_\phi$ can be neglected in Eq.~\eqref{eq:boltzmannR}, which can be integrated to relate temperature and scale factor by the usual conservation of comoving entropy,
\be
	g_S(T)\,T^3\,a^3 = {\rm~constant},
	\label{eq:entropyconservation}
\ee
where $g_S(T)$ and $g_*(T)$ are the number of effective degrees of freedom respectively in entropy and energy at temperature $T$, for which we use the parametrisation in Ref.~\cite{Borsanyi:2016ksw}.

However, Eq.~\eqref{eq:entropyconservation} is not guaranteed to hold ``during" the NSC period if the non-standard ``substance" decays into SM particles. The decay products will thermalise and release additional entropy. In such a modified scenario, the decrease in temperature with the scale factor would be slower than the standard $T\propto 1/a$ law in Eq.~\eqref{eq:entropyconservation}. Therefore, we write down the relation for entropy conservation in a NSC as
\be
	g_S(T)\,T^{3}\,a^{3\alpha} = {\rm~constant},
	 \label{aT_relation}
\ee
where $0 < \alpha \leq 1$ is a quantity that we use to parametrise the entropy injection and the subsequent decrease in the temperature decrease with the scale factor. Note that entropy conservation relates to $\alpha = 1$, for which Eq.~\eqref{aT_relation} returns the standard case in Eq.~\eqref{eq:entropyconservation}, while values of $\alpha>1$ would require a substance that efficiently sucks up energy from the SM bath, not contemplated in this study.

It proves useful to introduce a generic relation between time and scale factor,
\begin{equation}
	a \propto t^\beta, 
	\label{at_relation}
\end{equation}
where we assume the range $0 < \beta < 1$, corresponding to a cosmology in which the expansion is decelerating. Formally, $\beta=2/(3(1+w))$ where $w$ is the mean equation of state of the various fluids that are present at a certain time. During the NSC phase, the equation of state is well approximated with that of the substance dominating the energy density for times $t < t_{\rm RH}$, so $w\approx w_\phi$. Eq.~\eqref{at_relation} does not include the important case $w=-1, \beta \to \infty$ in which $a(t) \propto\exp(H\,t)$, which we do not treat explicitly. The Hubble rate during the modified cosmological epoch immediately follows from Eq.~\eqref{at_relation} as $H \equiv \dot{a}/a = \beta / t$. 

It is a simple exercise to solve Eqs.~\eqref{eq:boltzmannpsi} and~\eqref{eq:boltzmannR} at early times $t \lesssim 1/\Gamma_\phi$ and for a non-zero $\Gamma_\phi$, assuming that the radiation produced via the decay of the $\phi$ field dominates over any pre-existing radiation, see for example Refs.~\cite{Kolb:1998ki, Giudice:2000dp, Giudice:2000ex, Visinelli:2014qla, Visinelli:2016rhn, Freese:2017ace}. Under these conditions, using Eq.~\eqref{at_relation} and the definition $\rho_R\propto T^4$ (ignoring $g_*(T)$), we find an interesting relation between $\alpha$ and $\beta$~\cite{Giudice:2000ex}, 
\be
	\alpha = \frac{3}{8}(1+w_\phi) = \frac{1}{4\beta}.
\ee
To find the expression for $H(T)$ at early times, we combine Eqs.~\eqref{aT_relation} and~\eqref{at_relation} to obtain $t^{\alpha\beta}\propto 1/T$, so that the expression for $H(T)$ valid for $T \geq \TRH$ reads
\be
	H(T) \approx \HRH\,\left(\(\frac{g_S(T)}{g_S(\TRH)}\)^\frac{1}{3}\frac{T}{\TRH}\right)^{\frac{1}{\alpha\beta}}.
	\label{relation_H_T}
\ee

\subsection{Examples of modified cosmologies}

We consider explicitly three modified scenarios which are justified in known extensions of the Standard Model.

\begin{itemize}

\item Early matter decay domination (MD)

In this scenario, we take the early universe energy density to be dominated by a gas of NR massive particles (or a coherently oscillating NR field, which amounts to the same) for which we have $w_\phi=0,\beta = 2/3$. These particles have to decay in order for the standard cosmology to take place. This scenario has been used extensively to study low-reheat temperature (LRT) cosmologies~\cite{Dine:1982ah, Steinhardt:1983ia, Turner:1983, Scherrer:1985, Lazarides:1987zf, Lazarides:1990xp}, and it is famously problematic in the case of string moduli. One such tractable scenario embed in string theory is the large volume scenario~\cite{Balasubramanian:2005zx, Conlon:2005ki}, where a unique modulus field appears. The decay leads to a non-conservation of the entropy density $\alpha \neq 1$, in particular the model predicts $\alpha = 3/8$~\cite{Giudice:2000ex}. 

We present the evolution of $H$ and $a$ in this scenario in Fig.~\ref{fig:LTR} as red lines, obtained by solving numerically Eqs.~\eqref{eq:boltzmannR} and~\eqref{eq:boltzmannpsi} with $w_\phi=0$. As long as $\rho_\phi$ dominates over $\rho_R$ at some point, the decay rate $\Gamma_\phi$ fixes to a good extent the reheating temperature. In  Fig.~\ref{fig:LTR} we have adjusted $\Gamma_\phi$ to obtain $T_{\rm RH} \simeq 0.5\,$GeV and we have solved for different initial matter abundances (thicker lines have larger $\rho_\phi/\rho_R$ ratio). We distinguish three regimes for $H$: 1) At early times (high $T$) the radiation from the decaying matter field is subdominant with respect to a pre-existing radiation so effectively $\alpha=1$ and $H\propto a^{-3/2}\propto T^{3/2}$; 2) The radiation from the decaying field dominates over the pre-existing one. In MD cosmology, the Hubble rate $H$ decreases much faster with $T$ because $T$ is decreasing slower in time due to the radiation generated during the decay ($H\propto a^{-3/2}\propto T^{3/2\alpha}\sim T^{4}$); 3) The radiation-dominated period after the decay $H\propto T^{-2}$. If the initial matter abundance is much larger than the pre-existing radiation, only the second and third period will be relevant for us. We will mainly focus on this simple scenario (thick red line), for which indeed we reproduce $\alpha=3/8, \beta=2/3$ before reheating.

\begin{figure}[h!]
\includegraphics[width=0.45\textwidth]{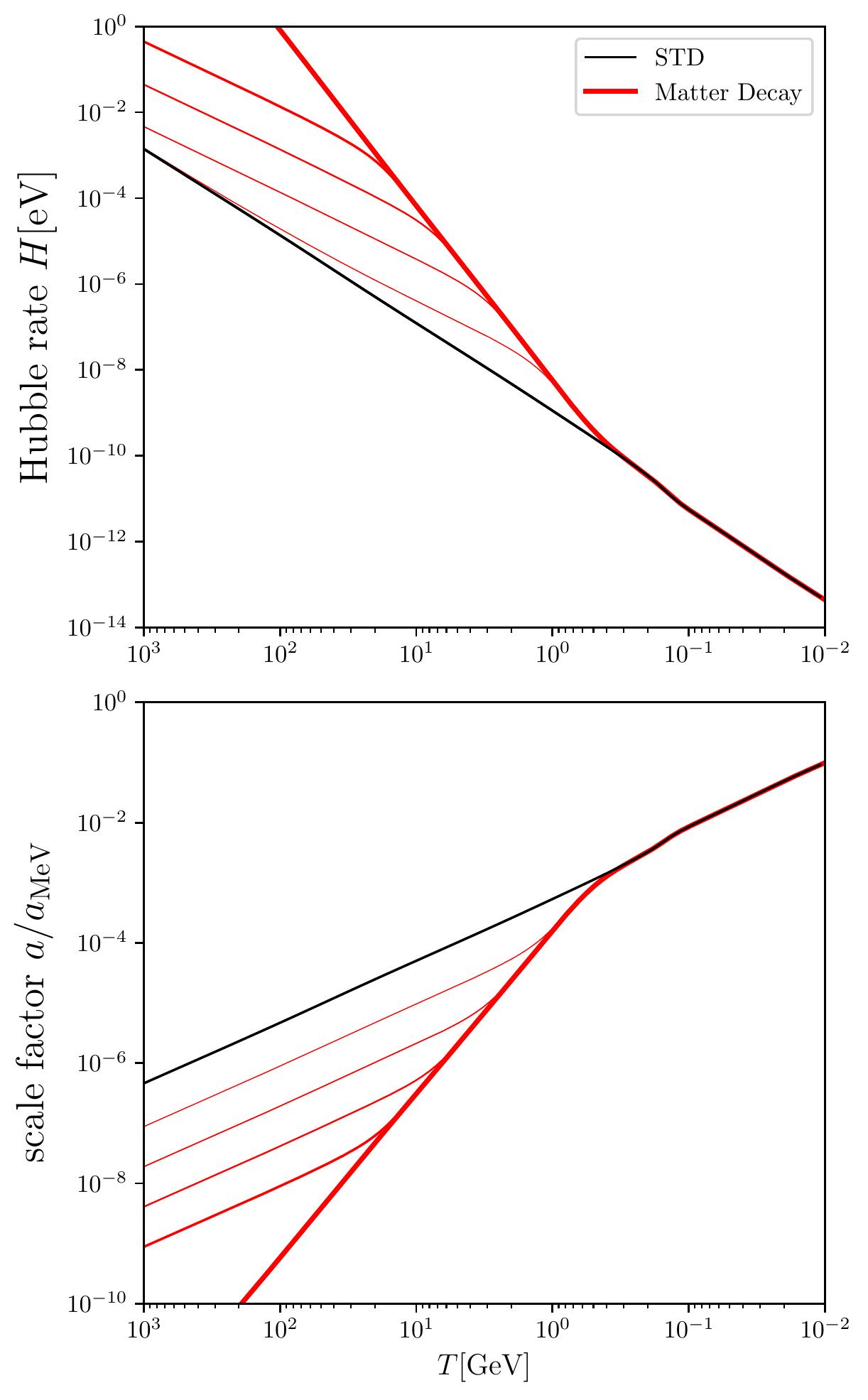}
	\caption{Evolution of Hubble rate and scale factor (normalised to the value at $T= 1\,$MeV) in Early matter domination (MD) cosmologies (red lines) compared to standard radiation domination (black). The red lines differ only in the initial matter/radiation fraction, which increases with increasing line-thickness. All the models share the same decay rate $\Gamma_\phi$, here set to obtain $T_{\rm RH} \sim 0.5\,$GeV. }
	\label{fig:LTR}
\end{figure}

\item Early Kination domination

In the Kination cosmology~\cite{Barrow:1982, Ford:1986sy, Spokoiny:1993kt, Joyce:1996cp, Salati:2002md, Profumo:2003hq}, the early universe expansion is driven by the kinetic energy of a scalar field. The field $\phi$ is a ``fast-rolling'' field with an equation of state $w_\phi=1$ (pressure $p_\phi$ equals the energy density $\rho_\phi$). The energy density of the field $\phi$ scales as $\rho_\phi \sim a^{-6}$, so that it redshifts faster than the radiation energy density $\rho_R \sim a^{-4}$ and becomes subdominant below $\TRH$. Therefore, during Kination domination we have $\beta = 1/3$. The thermal production of a WIMP during Kination has been discussed in Refs.~\cite{Salati:2002md, Profumo:2003hq, Pallis:2005hm, Pallis:2005bb, Gomez:2008qi, Lola:2009at, Lewicki:2016efe, Artymowski:2016tme, Redmond:2017tja, DEramo:2017gpl, Visinelli:2017qga, Redmond:2018xty}. 
Since the Kination field $\phi$ does not decay but it redshifts away, we have entropy conservation with $\alpha  =1$. 

The solution of Eqs.~\eqref{eq:boltzmannpsi} and~\eqref{eq:boltzmannR} with $w_\phi=1$ is shown in Fig. \ref{fig:KKK} as a blue line ($H\propto a^{-3}\propto T^{3}$). The reheating temperature can be adjusted by the initial value of $\rho_\phi$ for a given temperature (equivalently $\rho_R$). In Fig. \ref{fig:KKK} we have adjusted the initial value of $\rho_\phi$ to obtain $T_{\rm RH}\simeq 0.5\,$GeV.  

\begin{figure}[h!]
\includegraphics[width=0.45\textwidth]{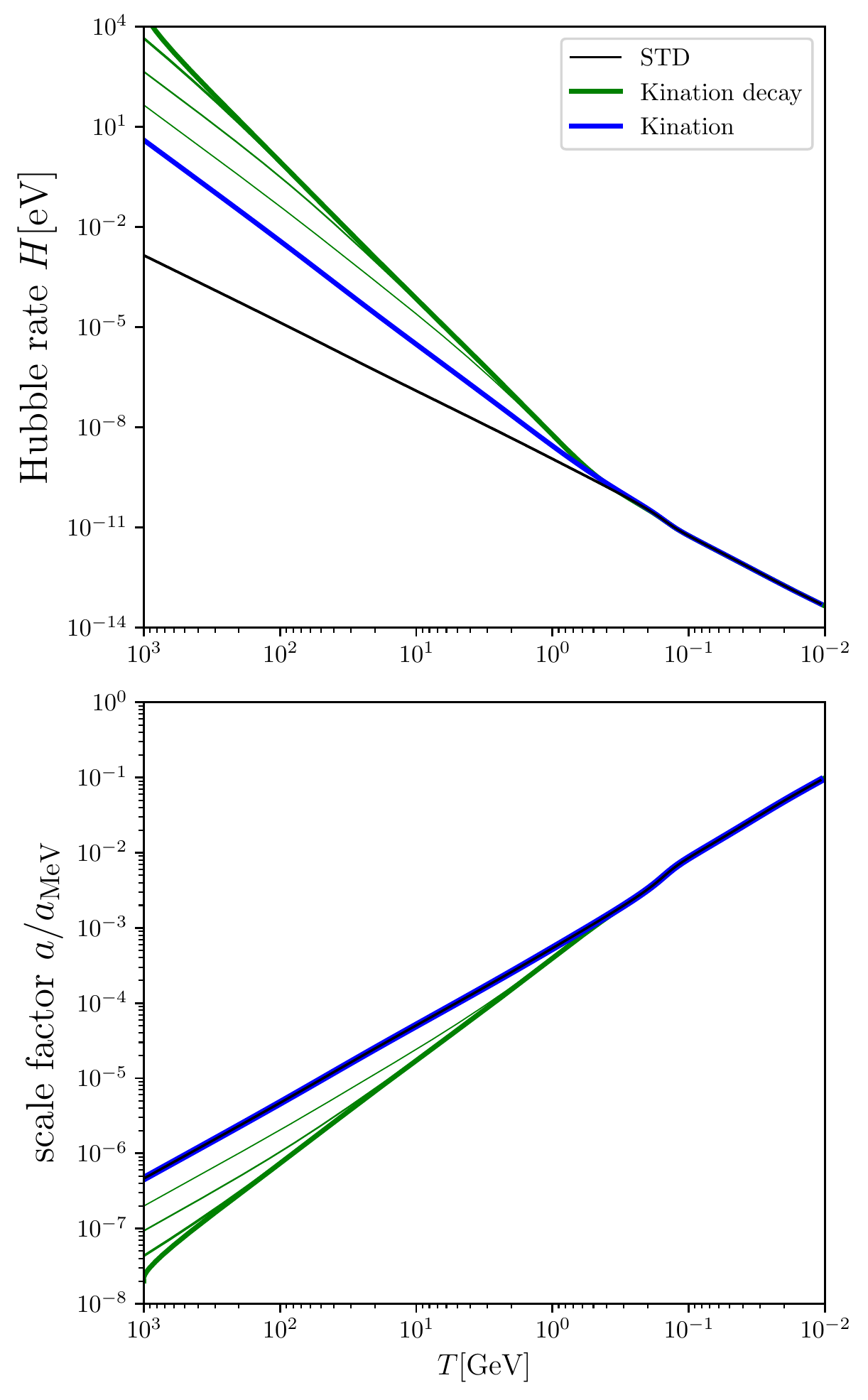}
	\caption{Evolution of Hubble rate and scale factor (normalised to the value at $T = 1\,$MeV) in Kination cosmologies (blue line) and decaying Kination (KD) (green lines) compared to standard radiation domination (black). The decaying Kination lines differ only in the initial Kination/radiation fraction, which increases with increasing line-thickness. All the models share the same decay rate $\Gamma_\phi$, here set to obtain $T_{\rm RH}\sim 0.5\,$GeV. In the non-decaying scenario, the initial abundance has been set to obtain the same $T_{\rm RH}$. }
	\label{fig:KKK}
\end{figure}

\item Kination decay (KD)

We also consider a decaying $\phi$ field as in Ref.~\cite{Visinelli:2017qga}. 
In this case, the reheating temperature is set by the abundance if the decay rate is very small and 
by the decay rate if the abundance is very large. 
The second makes the limiting case of interest here. 
In Fig. \ref{fig:KKK} we show cosmological histories for different initial Kination densities with respect to pre-existing radiation (increasing abundance follows the line thickness). As in the MD case, when the thermal bath is dominated by the pre-existing radiation, the Hubble rate $H$ follows a softer power law ($H\propto a^{-3}\propto T^3$) than when the radiation from the $\phi$-decay dominates\footnote{The fact that the latter coincides with the MD case is not a coincidence. For any decaying substance when its radiation dominates the temperature $H\propto a^{-1/\beta}\propto T^{1/\alpha\beta}\sim T^{4}$. } ($H\propto a^{-3}\propto T^{3/\alpha}=T^4$). Deep inside the KD period (thick green line in Fig.~\ref{fig:KKK}), one finds $\alpha = 3/4$ and $\beta = 1/3$. 
\end{itemize}

\subsection{The dark matter axion mass} \label{bound_axion_mass}

For a given NSC, there is one specific value of the axion mass (or equivalently of the axion decay constant $f_A$) for which axions account for all of the observed CDM ($\rho_{\rm CDM} \sim 9.72\times 10^{-48}{\rm\,GeV^4}$~\cite{Ade:2015xua}). Such a value of $m_A(T\to 0)$, here denoted as $m_{\rm CDM}$ and called the ``DM axion mass'', is in general a function of the parameters that define the NSC model ($T_{\rm RH}, \alpha, \beta$), as well as the $\at$ factor. We refer to the value in the standard cosmological scenario as $m_{\rm CDM}^\std$. Generally we denote by $m_A$ the value of the axion mass at zero temperature and we note explicitly the temperature as $m_A(T)$ when relevant.

In the remainder of the paper we want to assume for simplicity that axions account for all the CDM so we need to compute it for the different NSCs of interest.  
Technically, we compute these quantities as follows. Given a NSC model, equating $\bar \rho_A = \rho_{\rm CDM}$, with $\bar \rho_A$ as in Eq.~\eqref{eq:totalaxionenergydensity}, imposes a direct constraint on the value of the temperature $T_1$, so that using the definition in Eq.~\eqref{eq:111} we obtain the value of $f_A$ as a function of $\at$ for which the axion is the CDM particle. Since this calculations have been presented elsewhere~\cite{Abbott:1982af, Dine:1982ah, Preskill:1982cy}, here we only outline the key aspects that differentiate the computation in NSC models. For simplicity in the exposition we neglect $O(1)$ factors involving the slowly varying numbers of degrees of freedom, $g_{*}(T), g_{*S}(T)$. 

A technical point is that the topological susceptibility $\chi(T)$, and likewise the axion mass, depends on temperature. In the numerics of this paper we use the recent lattice calculation in the lattice from Ref.~\cite{Borsanyi:2016ksw}, but when considering analytical calculations a truncated power-law expression is useful~\cite{Gross:53.43}, as
\be
	\(\frac{m_A(T)}{m_A(0)}\)^2 = \frac{\chi(T)}{\chi(0)}= \begin{cases}
		1 & \hbox{for $T \leq \TL$},\\
		\left(\frac{\TL}{T}\right)^{2\gamma}, & \hbox{for $T \geq \TL$}.
	\end{cases}
	\label{axionmass}
\ee
where $T_\Lambda \simeq 160\,$MeV and $\gamma\sim 4$.

In general, the relevant sequence of events in the history of the axion field in a NSC reads
\be
	a_1 < a_{\rm RH} < a_\Lambda,
\ee
{\it i.e.}, first the axion field begins to oscillate during the NSC part of the early universe ($a_1$), then the universe converges to the standard radiation-dominated description ($a_{\rm RH}$), then the topological susceptibility saturates with $a_\Lambda = a(T_\Lambda)$. The latter turns out to be true for the ranges of parameters which are not yet excluded. 

The first important thing to note is that the present axion CDM energy density depends on the NSC parameters only through $T_1$, or equivalently $a_1$. For a given value $a_1$ of the scale factor at the onset of oscillations, the present energy density $\rho_0(a_1)$ is
\be
	\rho_0(a_1) \propto \frac{a_1^3}{T_1^\gamma} \propto a_1^{3+\gamma \alpha}, \quad\quad (a_1<a_{\rm RH}),
	\label{rho0a1}
\ee
where in the last equality we have used the parametrisation for the scale factor $a=a(T)$ in \mbox{Eq.~\eqref{aT_relation}} when the production of entropy is considered. Note that if the effect of a NSC is that of decreasing the scale factor $a_1$ at the onset of axion oscillations, then $T_1$ should also decrease in order to attain the observed CDM abundance $\rho_{\rm CDM}$ using Eq.~\eqref{rho0a1}.

\begin{figure}[t]
\includegraphics[width=0.45\textwidth]{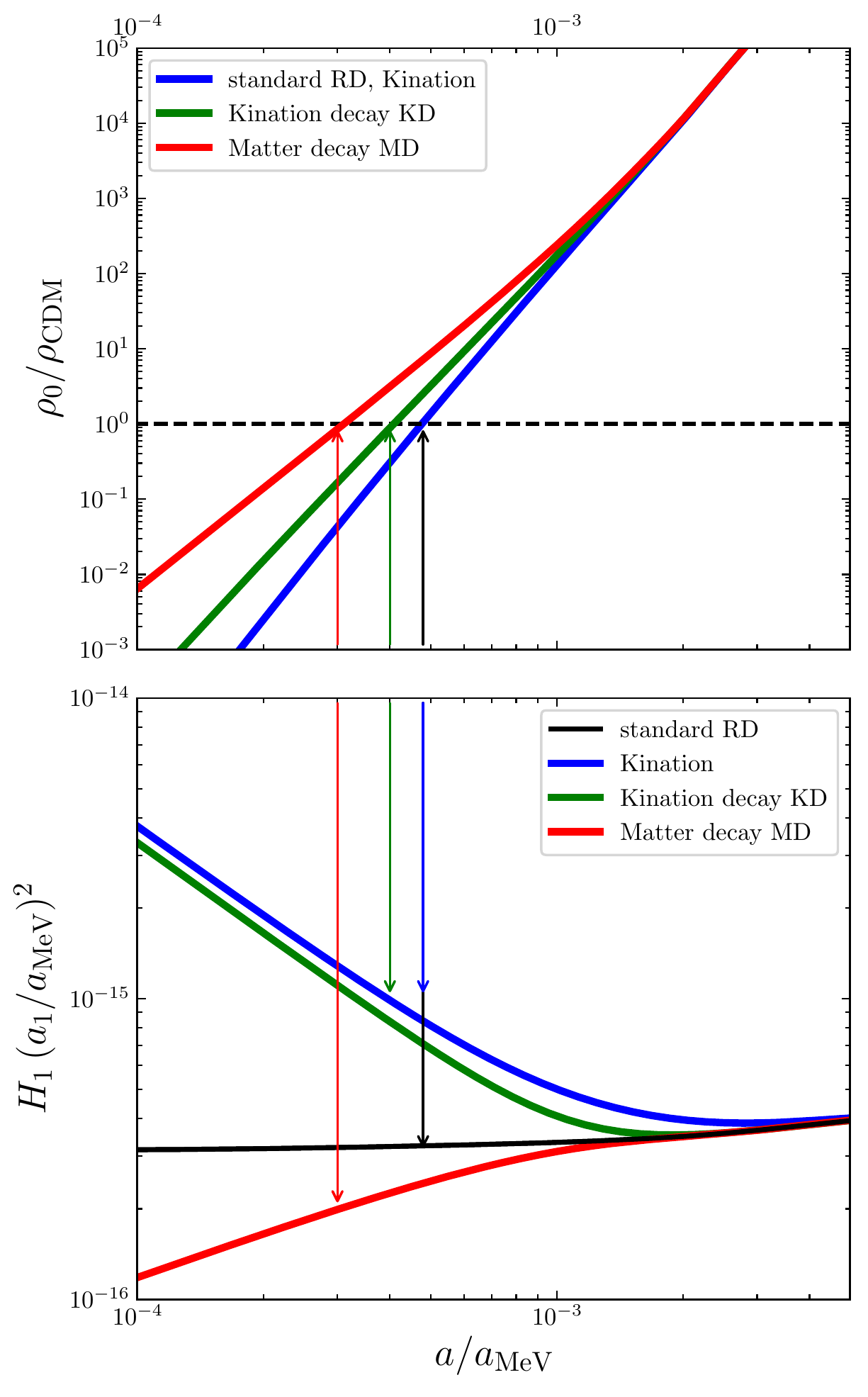}
	\caption{Top: Evolution of $\rho_0(a)$ in different non-standard cosmologies considered in the text for the same ``reheating" temperature $\TRH=0.5\,$GeV. Imposing $\rho_0=\RCDM$ at scale factor $a_1$ we find $a_1\leq a_1^\std$. Bottom: The Hubble expansion rate, $\times a^2$, captures the main trend of the DM axion mass, \eqref{mmm}.}
	\label{fig:ADMM}
\end{figure}

\emph{We first focus on comparing different NSCs with the same $\TRH$.} Later we will discuss the effects of $\TRH$ for a given cosmology. In Fig.~\ref{fig:ADMM} (top panel), we plot a scheme of the evolution of $\rho_0(a_1)$ as a function of $a_1 = a$ in the standard radiation dominated cosmology (RD), Kination (K), Kination decay (KD) and matter decay (MD) scenarios. We also show a horizontal line denoting the observed value that we are interested to reproduce. The point at which $\rho_0(a)$ intersects $\RCDM$ corresponds to $a_1$ in the different cosmologies and yields the value $\bar\rho_A$ in Eq.~\eqref{eq:totalaxionenergydensity}. By defining the ratio of $\rho_0(\aRH)$ to $\RCDM$ as $Q \equiv \rho_0(\aRH)/\RCDM$ and using Eq.~\eqref{rho0a1}, we find 
\be
	a_1 = \frac{\aRH}{Q^\frac{1}{3+\gamma\alpha}}, 
	\label{a1}
\ee 
so that the comparison with the case in which entropy is conserved $\alpha=1$, like in the standard cosmology, gives
\be 
	\frac{a_1}{a_1^\std}= Q^{\frac{1}{3+\gamma}-\frac{1}{3+\gamma\alpha}}.
	\label{a1astd}
\ee
We obtain that the scale factor at which axions become CDM \emph{depends only on $\alpha$ but not on $\beta$. } The dependence on $\TRH$ is hidden in the definition of $Q$ and will be dealt with later. 

If entropy is conserved in the NSC ($\alpha=1$), oscillations commence at the same value of the scale factor as in standard radiation dominated history, $a_1=a_1^{\std}$. In this case, we should also have $T_1=T_1^\std$ in order to respect the CDM constraint. In our small collection of NSCs, this is the case of red-shifting Kination cosmology.

By contrast, if during the NSC period there has been an extra injection of entropy, the temperature redshifts slower than $1/a$ ($\alpha<1$), the topological susceptibility increases more slowly with $a$, therefore we have to get an earlier onset of the axion oscillations to get the same final CDM, {\it i.e.} $a_1<a_1^\std$. Note that the value of $a_1$ in Eq.~\eqref{a1astd} decreases with decreasing $\alpha$, therefore we have 
\be
	a_1^{\rm MD}<a_1^{\rm KD}<a_1^{K}=a_1^\std,
	\label{eq:hierarchy}
\ee 
because $\alpha=3/8$ during MD, $\alpha=3/4$ during KD, and $\alpha = 1$ during both Kination and the standard cosmology.

Once the value for $a_1$ has been obtained, the temperature at which axion oscillations commence, $T_1$, is also computed. For this, just consider the ratio of Eq.~\eqref{rho0a1} for the NSC and standard cosmology giving the same $\RCDM$, 
\be
	\frac{T_1}{T_1^\std} = \(\frac{a_1}{a_1^\std}\)^{3/\gamma}, 
	\label{a1T1}
\ee
which implies the relation between temperatures
\be
	T_1^{\rm MD}<T_1^{\rm KD}<T_1^{K}=T_1^\std . 
\ee
Writing Eq.~\eqref{eq:111} in the form $3H_1 = m_{\rm CDM} \({\TL}/{T_1}\)^\gamma$ and taking ratios of such an expression when evaluated for a NSC and the standard scenarios gives
\be
	\frac{m_{\rm CDM}}{m_{\rm CDM}^\std} = \frac{H_1}{H_1^\std}\(\frac{T_1}{T_1^\std}\)^\gamma =\frac{H_1}{H_1^\std}\(\frac{a_1}{a_1^\std}\)^3 ,  
	\label{mmm}
\ee
where in the last step we used Eq.~\eqref{a1T1}. Therefore, the DM axion mass depends on the particular product $H_1 a_1^3$. 
To proceed further we note that the Hubble expansion rate as a function of $a$ during the NSC period is\footnote{In this formula there is a small correction due to our definition of $\tRH$, Eq.~\eqref{eq:Hubble_reheat}, which results in a difference in the two Hubble rates of the order of $ \approx 10\%$ when $a\ll \aRH$. This factor does not affect our qualitative discussion, but appears in Fig.~\ref{fig:ADMM}. For this reason we use $\simeq$ instead of the equality in the rest of the Section.}, 
\be
	H(a) \simeq \HRH\(\frac{\aRH}{a}\)^{1/\beta},  \quad  \quad  (a \lesssim \aRH)\,. 
	\label{HHH}
\ee
In standard RD cosmology with $\beta = 1/2$ we get $H_1 a_1^2 = $constant, if we neglect changes due to $g_*(T)$. To compare NSCs with the standard RD case we show the combination $H_1 a_1^2$ for the different cosmologies considered in the bottom panel of Fig.~\ref{fig:ADMM}. 

In the case of Kination cosmology, for which $a_1^{\rm K} = a_1^\std$, the value of $H_1$ is larger than $H_1^\std$ due to the Kination field whose energy dominates the expansion. We conclude from Eq.~\eqref{mmm} that $m_{\rm CDM}^{\rm K}/m_{\rm CDM}^\std = H_1^{\rm K}/H_1^\std > 1$, {\it i.e.} the DM axion mass in the Kination cosmology is generally \emph{larger} than in standard scenario.

In the case of the KD scenario, $a_1^{\rm KD} < a_1^\std$ and obviously $H_1^{\rm KD} > H_1^\std$ so the DM axion mass is not immediately clear. Indeed, a very interesting situation happens. Since $H_1\propto 1/a_1^3$ during Kination cosmology, both factors of $a_1$ in Eq.~\eqref{mmm} cancel out and the result becomes independent of the particular value of $a_1$ and equal to $m_{\rm CDM}^{\rm KD}/m_{\rm CDM}^\std =\aRH/a_1^\std$, which turns out to be $H_1^{\rm K}/H_1^\std$, similarly to what we discussed for the case of the Kination cosmology without the decay term.

In the case of matter decay (MD) we have again $a_1^{\rm MD} < a_1^\std$ so again the DM axion mass depends on the balance between the $H_1$ and $a_1^3$ factors. In this case, note that we have $H_1^{\rm MD}(a^{\rm MD}_1)^2 < H_1^\std (a_1^\std)^2$ because matter density redshifts slower than radiation. Therefore we conclude that the DM axion mass in the early decaying matter cosmology is generally \emph{smaller} than in the standard cosmology, $m_{\rm CDM}^{\rm MD}/m_{\rm CDM}^\std <1$. 

Combining Eqs.~\eqref{mmm} and~\eqref{HHH}, we obtain the dependence of the DM axion mass on the scale factor,
\be
	\frac{m_{\rm CDM}}{m_{\rm CDM}^\std} \simeq \frac{\aRH}{a_1^\std}\(\frac{a_1}{\aRH}\)^{3-\frac{1}{\beta}} \,, 
	\label{rata}
\ee
which makes very clear that for Kination cosmology ($\beta=1/3$), the DM axion mass only depends on the scale factor of reheating, it is larger or equal than the standard value (because $\aRH/a_1^\std\geq 1$). Also, since the main dependence on $a_1$ cancels out, the DM mass can only depend mildly on whether the Kination field is decaying or not.

Let us now study the dependence of the value of the DM axion mass on $\TRH$. First, let us note that using standard cosmology (for instance) we have
\be
	Q = \(\frac{T_1^\std}{\TRH}\)^{3+\gamma}\,,
\ee 
so that we replace Eq.~\eqref{a1} into Eq.~\eqref{rata} to write 
\be
	\frac{m_{\rm CDM}}{m_{\rm CDM}^\std} \simeq Q^{\frac{1}{3+\gamma} - \frac{3-1/\beta}{3+\gamma\alpha}} =\(\frac{\TRH}{T_1^{\rm std}}\)^{-1+\frac{3\beta-1}{\beta}\frac{3+\gamma}{3+\gamma\alpha}}. 
	\label{gorgi}
\ee
Therefore, for the NSC scenarios here considered, we obtain
\bea
	m_{\rm CDM}^{\rm K} \simeq m_{\rm CDM}^{\rm KD} &\simeq& m_{\rm CDM}^\std \(\frac{T_1^\std}{\TRH}\),\label{mcdmKK} \\
	m_{\rm CDM}^{\rm MD} &\simeq& m^\std_{\rm CDM} \(\frac{\TRH}{T_1^\std}\)^{\frac{4+3\gamma}{8+\gamma}},\label{mcdmMD}
\eea 
where the last exponent is $\sim 4/3$ for $\gamma \sim 4$. The chosen NSCs have the virtue of representing scenarios in which the DM axion mass differs from the standard case in opposite ways, since $m_{\rm CDM}^{\rm K}$ and $m_{\rm CDM}^{\rm KD}$ are generally larger than $m_{\rm CDM}^\std$ while $m_{\rm CDM}^{\rm MD}$ is generally smaller than $m_{\rm CDM}^\std$.

Eq.~\eqref{mcdmKK} shows that the mass of the axion CDM in both Kination and Kination Decay models is approximately the same. This might seem surprising because of the different relation relating the scale factor and temperature in the two cosmologies. However, the smaller value of $T_1$(or $a_1$) in the KD scenario (with respect to the K scenario) means that the axion mass at $T_1$ is larger (than in K), but $H_1$ is also larger by the same factor, so that the equality $3H_1  = m_{\rm CDM} (T_\Lambda/T_1)^\gamma$ is satisfied by the same value of $m_{\rm CDM}$. As explained, this is a bit coincidental and happens only for Kination where $1/\beta=3$, essentially because $H$ during Kination redshifts with the scale factor $a$ as CDM density.  This coincidence is obtained by modeling the evolution of $a_1,T_1,H$ by broken power-laws and will be corrected by small O(1) factors when the continuous evolution is taken into account, factors of $g_*, g_S$ are incorporated and by the fact that $H(a)$ differs by O(10\%)  between K and KD scenarios due to our definition of $\tRH$.

\begin{figure}[h!]
\includegraphics[width=0.5\textwidth]{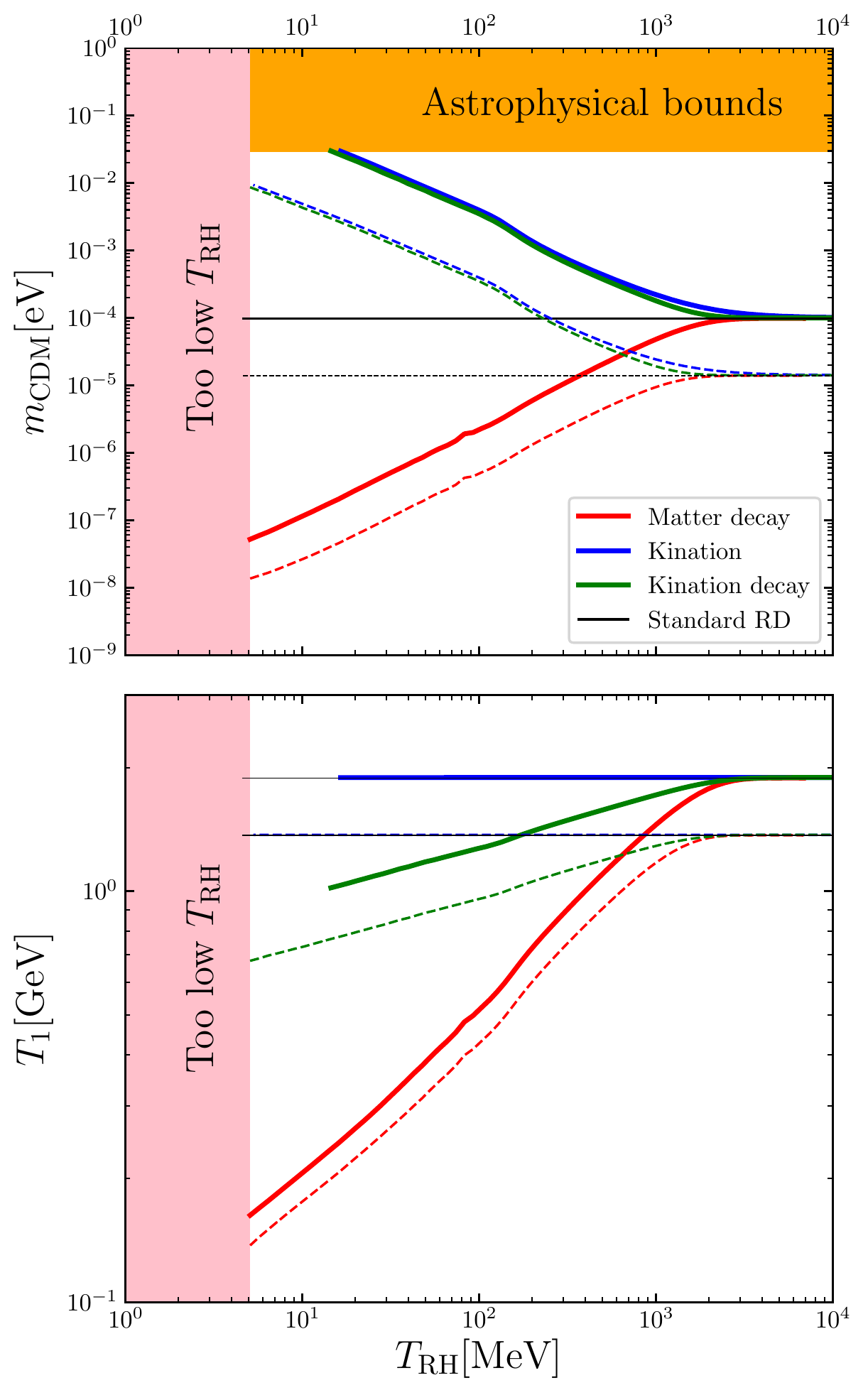}
	\caption{Top panel: The QCD DM axion mass as a function of the reheat temperature for different cosmological models. Black line: (standard) radiation-dominated cosmology. Red line: matter decay (MD) cosmology. Blue line: Kination. Green line: Kination with a decaying field (KD). Bottom panel: the temperature $T_1$ at which the axion field begins coherent oscillations, as a function of the reheat temperature. Solid and dashed lines assume $\at=10$ or $\at = 1$, respectively.}
	\label{fig:massCDM}
\end{figure}

We have performed full numerical calculations, adequately including all the relevant temperature-dependent functions like $g_{*}(T)$, $g_{*S}(T)$, the susceptibility $\chi(T)$ from Ref.~\cite{Borsanyi:2016ksw} and the numerical results for the cosmological evolution of $H(a),T(a)$, etc., computed in the previous section. In particular, we have solved numerically for $T_1$ and $m_{\rm CDM}$ as a function of the temperature $T_{\rm RH}$ in the different NSCs. Our results are shown in Fig.~\ref{fig:massCDM}. 

The bottom panel of Fig.~\ref{fig:massCDM} shows the value of the temperature $T_1$ required for having the totality of the CDM in axions. The black thin line represents the value of the DM axion mass in the standard cosmology. As expected, results in the standard and in the Kination cosmologies superimpose because entropy is conserved in both scenarios and $a_1 \propto 1/T_1$ is the same. The cosmologies with a decaying field, KD (green) and MD (red) require smaller values of $T_1$ (and of $a_1$) to obtain the same CDM abundance despite the larger entropy production and subsequent CDM dilution. We have coloured in pink the region where the reheating temperature is smaller than $\sim 5\,$MeV and is thus excluded by BBN considerations~\cite{Hannestad:2004px}. It is an interesting fact that this constraint limits the lowest value of $T_1$ to be $\sim \TL$ (in the MD scenario) so that our previous analytical considerations where we assumed $T_1>\TL$ are justified. For illustration, we show results for two values of the topological defects contributions $\at = 10$ and $\at = 1$ in solid and dashed lines, respectively. The difference is a minimal shift in $T_1$, less severe in the MD case.

The upper panel of Fig.~\ref{fig:massCDM} shows the DM axion mass. The black thin line sets the value in the standard scenario. 
Again, we show results for two values of $\at = 10$ and $\at = 1$ in solid and dashed lines, respectively. We note as usual that larger $\at$ requires a larger $m_{\rm CDM}$ following a roughly linear dependence. As expected, Kination and KD cosmologies require a similar DM axion mass because of their particular expansion rate, $\beta=1/3$ that cancels the dependence on $a_1$. This does not cancel the dependence with the reheating temperature $\sim 1/\TRH$ predicted in Eq.~\eqref{mcdmKK}. In the Kination scenario, the DM mass becomes so large that it can conflict with the astrophysical bounds at $m_A \gtrsim 30\,$meV~\cite{RAFFELT1986402, Raffelt2008, Viaux:2013lha} (orange region) so we have interrupted the line at this approximate mass. This range of values is particularly interesting as axions could explain several astrophysical anomalies~\cite{Giannotti:2017hny} and be detected by IAXO~\cite{Vogel2015}. However, this statement depends on the assumed value of $\at$. Using $\at = 1$, one hits the BBN constraint in Eq.~\eqref{reheatingconstraint} before the astrophysical bounds. The value of $\mCDM$ in a Kination model can change up to two orders of magnitude and is only limited indirectly by the astrophysical bounds or BBN. 

The matter decay case is shown in red and is well described by $m_{\rm CDM}\propto \TRH^{4/3}$. The axion CDM mass can be almost three orders of magnitude smaller than the standard value before conflicting BBN. In general, the full numerical result agrees very well with the analytical trends reviewed there. The only apparent differences are attributable mostly to the matching of cosmologies around $\TRH$ and to the changes in $g_*(T)$ and $g_S(T)$ to a lesser extent.

\subsubsection{Effects of pre-existing radiation in MD and KD}

In principle, it is straightforward to extend our results to the case in which the pre-existing radiation before the decay of the matter or Kination field $\phi$ dominates the temperature during the period when axions become CDM. This amounts to consider a period of $\alpha=1$ during the NSC period. The physical idea is that the pre-existing radiation redshifts as $\propto a^{-4}$ so $T \propto 1/a$ and $\alpha=1$.  Let us denote by $T_r, a_r$ the temperature and scale factor at which the radiation resulting from the decay of the $\phi$ field catches up with the pre-existing amount. A first modification is that $\rho_0 = \rho_0(a)$ increases as $a^{3+\gamma}$ before $a_r$, which is steeper than the $\propto a^{3+\gamma\alpha}$ behaviour shown in Fig.~\ref{fig:ADMM} (top) This makes the required value of $a_1$ to lie between the value of $a_1$ calculated in a NSC model (here labelled $a^{\rm rd}_1$) and $a_1^\std=a_1(\alpha=1)$. Letting $a_r$ vary within the range $(a_1^{\rm rd},\aRH)$ yields $a_1$ in the range $(a_1^{\rm rd}, a_1^\std)$. At this stage, Eq.~\eqref{a1T1} is still valid, so that the temperature $T_1$ lies in the interval $(T_1^{\rm rd}, T_1^\std)$, corresponding to the area between a given NSC line and the STD line in Fig.~\ref{fig:massCDM} (bottom). The derivation of the DM axion mass $m_{\rm CDM}$ is a bit more involved for a similarly simple and not surprising outcome: depending on the values of the parameters $a_r$ and $T_r$, the DM axion mass can take any value between that of a given NSC line shown in Fig.~\ref{fig:massCDM} (top) and the standard case $m_{\rm CDM}^\std$. For completeness, the expression equivalent to Eq.~\eqref{gorgi} reads 
\be
	\frac{m_{\rm CDM}}{m_{\rm CDM}^\std} = \(\frac{\TRH}{T_1^{\rm std}}\)^{2-\frac{1}{\beta}} \(\frac{\TRH}{T_r}\)^{\frac{\gamma}{\alpha}\frac{1-\alpha}{3+\gamma}\frac{3\beta-1}{\beta}}.  
\ee

\section{Axion miniclusters} \label{miniclusters_analytic}

Around the temperature $T_1$ of the onset of oscillations, the axion field is highly inhomogeneous. In the standard radiation dominated scenario, the density field becomes frozen in comoving coordinates around that time~\cite{Hogan:1988mp}. At the same time, axions are rapidly acquiring a non-zero mass, faster than the redshift of their momentum with the expansion of the universe so that they become non-relativistic and cannot further free-stream. Because of this, the correlation length of the axion DM density field is of the order of the causal horizon at $T_1$, {\it i.e.} $\sim 1/H_1$, and freezes in comoving coordinates. This picture, sketched by the seminal studies~\cite{Kolb:1993zz, Kolb:1993hw, Kolb:1994fi} has been recently confirmed in large scale numerical computations including cosmic strings~\cite{Vaquero:2018tib}. Although the results show a rich structure of even larger inhomogeneities at sub-horizon scales due to the dynamics of cosmic-strings, domain walls and axitons, these substructures encapsulate a relatively small amount of the DM~\cite{Vaquero:2018tib} so we will ignore it henceforth.  
In this section we derive the properties of the miniclusters in non-standard cosmologies. We begin by a brief description of the standard case. 

\subsection{Standard cosmology}

\emph{Density of the minicluster.-}
The density inhomogeneities separate out from the cosmic expansion as gravitationally bound miniclusters around matter-radiation equality~\cite{Kolb:1993zz}, happening around a redshift is $z_{\rm eq}=a_0/a_{\rm eq}-1\simeq 3360$ and plasma temperature $\Teq=T_0 a_0/a_{\rm eq}\simeq 0.8$ eV, much after our NSC has converged to a standard cosmology.  
We parametrise an overdensity in the axion CDM density as~\cite{Kolb:1994fi}
\begin{equation}
	\Phi = \frac{\rho_c - \bar\rho_A}{\bar\rho_A},
	\label{eq:def_Phi}
\end{equation}
where $\rho_c$ is the local density of the minicluster and $\bar \rho_A$ is the mean axion energy density. Since the energy density in axion CDM divided by that of radiation scales as $\bar \rho_A/\rho_R\propto a$, an overdense region with $\Phi>0$ enters into matter-domination when the scale factor is $a_{\rm leq}= a_{\rm eq}/\Phi$. The overdensity grows linearly with the scale factor until it becomes of order unity, when collapses and forms a gravitationally bound virialised object. Since $\Phi$ is typically of order unity~\cite{Vaquero:2018tib}, miniclusters collapse typically around matter radiation equality. Note that values $\Phi\gg 1$ are also possible, although these regions tend to be smaller than the causal horizon at $T_1$ and thus correspond to smaller, much less massive miniclusters, see below. The energy density of the minicluster at the moment of collapse is
\bea
	\rho_c(T_{\rm collapse}) &=& (1+\Phi)\rho_{a}(T_{\rm collapse}) = \nonumber\\
    	&=& (1+\Phi)\(\frac{a_{\rm eq}}{a_{\rm collapse}}\)^3 \rho_A^{\rm eq} = \nonumber\\
    	&=& (1+\Phi)\Phi^3 \rho_A^{\rm eq},
	\label{eq:parametrize}
\eea
where $\rho_A^{\rm eq}$ is the axion energy density at matter-radiation equality. A detailed calculation that follows from the dynamics of the spherical collapse and further virialisation of the minicluster obtains an extra factor of 140~\cite{Kolb:1993zz, Kolb:1994fi}, so the expression we use in place of Eq.~\eqref{eq:parametrize} for the energy density of the minicluster is
\be
	\rho_c = 140\(1+\Phi\)\,\Phi^3\rho_A^{\rm eq}.
	\label{density_clusters}
\ee
Recall that we will be assuming that axions account for the entirety of the CDM, so that we can write $\rho_R^{\rm eq}=\rho_m^{\rm eq} = \rho_A^{\rm eq}(1+\Omega_b/\Omega_{\rm CDM})$, where $\Omega_b/\Omega_{\rm CDM} \sim 0.19$ is the ratio of baryonic matter to CDM. Using $\rho_A^{\rm eq}=\rho_{\rm CDM}(1+z_{\rm eq})^3$, this gives
\be
	\rho_c\sim 6.7 \times  10^6  (1+\Phi)\Phi^3\frac{\rm GeV}{\rm cm^3}.
	\label{rhoc}
\ee
The cores of miniclusters have the typical energy density of either matter or radiation components at matter-radiation equality. 

\emph{Mass of the minicluster.-}
The mass of the axion minicluster is given by the dark matter mass enclosed within the overdense region at about the onset of oscillations~\cite{Hogan:1988mp, Kolb:1993zz}. We consider a spherical region of physical radius $R_1$ at temperature $T_1$. According to the simulations in Ref.~\cite{Vaquero:2018tib}, the typical value of $R_1$ is of the order of $\sim 1/H_1$, although the temperature $T_1$ is better defined slightly differently than in Eq.~\eqref{eq:111}, that is without the factor of three which is useful to discuss the DM abundance, as $H_1=H(T_1)=m_A(T_1)$. Due to the large value of the exponent $\gamma$ describing the temperature-dependence of the axion mass, this modification amounts to only a $\sim 10\%$ differences in the values of $T_1$ and $a_1$. Here, we parametrise deviations from the canonical size as $R_1=r/H_1$, with $r =\mathcal{O}(1)$. In terms of the present CDM density, we have
\begin{equation}
	M_c^{\rm std} \!=\! \frac{4\pi}{3} (1+\Phi) \rho_1 R_1^3 \!=\! \frac{4\pi}{3}(1+\Phi)\rho_{\rm CDM}\!\(\frac{r}{H_1}\)^3\!\!\left(\frac{a_0}{a_1}\right)^{3},
	\label{mass_cluster}
\end{equation}
where $\rho_1 \equiv \at\rho_A^{\rm mis}(T_1) = \at\,m_A(T_0)\,n_A^{\rm mis}(T_1)$, with $\rho_A^{\rm mis}$ defined in Eq.~\eqref{eq:totalaxionenergydensity} and $n_A^{\rm mis}(T_1)$ defined in Eq.~\eqref{number_density_onset} below. Using the numerical calculations in Ref.~\cite{Vaquero:2018tib} to fix the quantity $a_0/a_1H_1\simeq 1.1 \times 10^{17} (50\mu {\rm eV}/m_A)^{0.1712}\,$cm, we find
\be
	M_c^{\rm std} =  6.6\times 10^{-12} (1+\Phi)r^3\(\frac{50 \mu\rm eV}{m_A}\)^{0.500} M_\odot,
	\label{mass_cluster_numeric}
\ee
where again we are assuming that $\at$ is chosen so that cold axions explain all of the CDM observed.

\emph{Radius of the minicluster.-} Given the results for the mass and the density of the minicluster, one estimates the radius of a minicluster as
\be
	R_c^{\rm std} \!=\! \(\frac{3M_c}{4\pi \rho_c}\)^{1/3} \!=\! 
	\frac{r}{(140)^{1/3}\Phi}\,\frac{\aeq}{ H_1 a_1},
     \label{size_minicluster}
\ee
which for our estimates in Eqs.~\eqref{rhoc} and~\eqref{mass_cluster_numeric} gives 
\be
	R_c^{\rm std} = 6.4\times 10^{12} {\rm cm} \frac{r}{\Phi} \(\frac{50 \mu\rm eV}{m_A}\)^{0.1668},
	\label{size_std}
\ee
which corresponds to about one astronomical unit. 

\subsection{Miniclusters in non-standard cosmologies}

The properties of axion miniclusters like its mass, its size, and the velocity dispersion, are modified in a number of ways by the presence of a NSC period. The most important and obvious way is that the correlation length of the axion DM field $\sim 1/H_1$ when axions become NR and thus DM can be totally different from the standard radiation-dominated scenario.

A second potentially relevant effect is a non-trivial evolution of axion overdensities during the NSC. Indeed, it is well known that during radiation domination, CDM perturbations only suffer a negligible logarithmical growth with the scale factor, but during a period of matter domination perturbations can grow linearly and produce already some collapses.

Another potential issue is that the relation $H_1\simeq m(T_1)$ defining the moment when axions become DM and the correlation length which gives us the size of the typical minicluster would suffer O(1) corrections in a NSC. A similar effect is the expected different values of $\at$ in different NSCs, due for instance to the O(1) differences of the string-network density, see Ref.~\cite{Yamaguchi:1999dy} for an example in matter domination and the generalisation in Ref.~\cite{Visinelli:2009kt}. Since the two effects discussed before vary the minicluster properties by several orders of magnitude, while these latter ones impact the energy density only by a factor of order one, we will not discuss this issue further here. 

\subsubsection{Impact of the horizon size at $T_1$} \label{Impact of the horizon size}

The implications of the change of $H_1$ and $a_1$ in NSCs to the properties of miniclusters are relatively trivial to study. First of all, we have already prepared the stage in the previous section when we studied the DM axion mass. 
Indeed, we showed in Eq.~\eqref{mmm}  how the DM mass in a NSC is proportional to $H_1 a_1^3$ (with an extra $a_1$) and how 
this quantity compares in the different NSCs. Moreover, we have performed the numerical computations and analytical estimates of $H_1, a_1$ as a function of the NSC parameters and $\TRH$. 
Therefore, assuming that the overdensities become frozen when $H_1=m_A(T_1)$ we can use the same rationale as in the standard case to compute the density, mass and radius of the minicluster. We will drop the assumption on the next subsection but we advance that it is actually quite reasonable. 

The typical minicluster density derived in Eqs.~\eqref{eq:parametrize},~\eqref{density_clusters}, and~\eqref{rhoc} \emph{does not depend at all on the NSC}. These equations only depend on the assumption that axions account for all the CDM and parameterise the effects of an early/late collapse with an overdensity parameter~$\Phi$. 

\begin{figure}[t!]
	\includegraphics[width=0.5\textwidth]{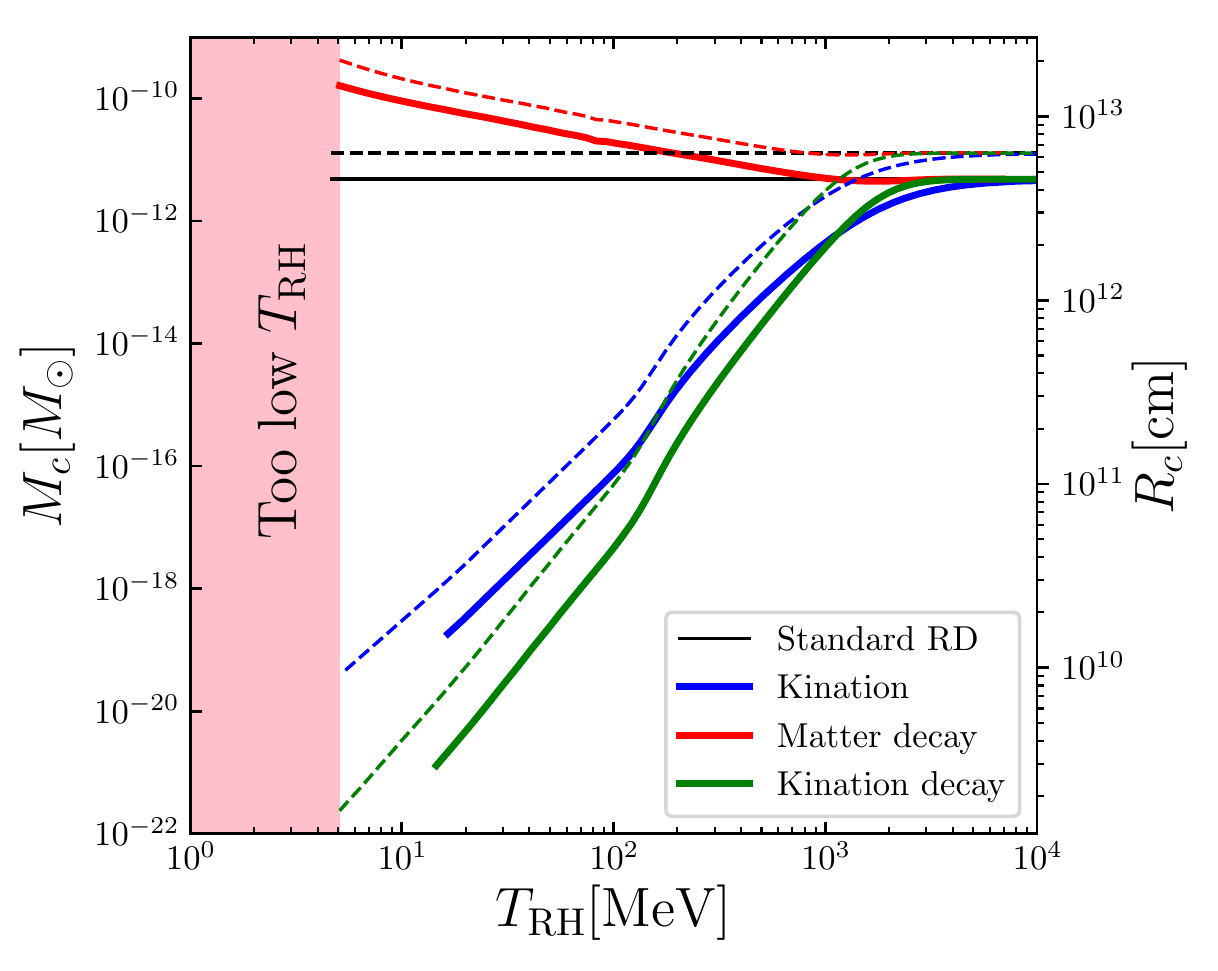}
	\caption{The mass (left vertical axis) of a typical axion minicluster ($r=1,\Phi=1$) in units of solar masses as a function of the reheat temperature for different cosmological models before nucleosynthesis. We also report the radius of the minicluster in cm (vertical right axis). We have assumed that 100\% of the CDM is in axions and $\Phi = 1$. Each value of the reheating temperature requires a different axion mass from Fig.~\ref{fig:massCDM}. Solid and dashed lines assume $\at=10$ or $\at = 1$, respectively.}
	\label{fig:miniclustermassradius}
\end{figure}

The typical minicluster mass computed in Eq.~\eqref{mass_cluster} also depends on the axion CDM assumption but involves the typical size of an inhomogeneity at $T_1$, given by a radius $R_1\sim H_1$. We note that the mass and the radius of the minicluster depend on the quantity $1/\(H_1a_1\)$. We have seen that, insisting on the axion CDM hypothesis, the value of $a_1$ in the NSC is smaller than the standard value unless no extra entropy is generated ($\alpha = 1$), see Eq.~\eqref{eq:hierarchy}. This effect points towards more massive miniclusters due to a larger DM density $\rho_A\propto \rho_{\rm CDM}/a_1^3$. However, in our NSCs, the causal horizon at $a_1$ is also smaller than the standard results, so that $H_1$ is larger, due to the presence of extra energy and the value of $a_1$ computed at earlier times. It is not straightforward to anticipate which of the two effects is the prominent one, due to the complexity in the expressions brought in by assuming that the axion is the CDM particle. Using the numerical results developed in the previous section we have computed the values of the typical minicluster mass $M_c$ and radius $R_c$ in the NSCs considered, as shown in Fig.~\ref{fig:miniclustermassradius}.

Notice that $M_c\propto 1/(H_1 a_1)^3$, so that we use the expressions for $H$ in Eq.~\eqref{HHH} and for $a_1$ in Eq.~\eqref{a1} to derive a simple analytical formula for the mass of a minicluster. Neglecting the variation in the number of degrees of freedom as in the previous section, we obtain
\be
	M_c = M_c^{\rm std}\,\left(\frac{\TRH}{T_1^{\rm std}}\right)^{3\(\frac{1-\beta}{\beta}\frac{3+\gamma}{3+\alpha\gamma}-1\)}.
	\label{mcmass}
\ee
In the NSCs, the most important contributions in the exponent turn out to come from the term $(1-\beta)/\beta$, which varies from $1/2$ to $2$ respectively for MD or Kination models, while the factor $(3+\gamma)/(3+\alpha\gamma)$ remains reasonably close to unity. The origin of this complicated exponent is nevertheless clear. Recall that $H_1\propto (\aRH/a_1)^{1/\beta}$ during the NSC epoch and therefore $1/(H_1 a_1)\propto (a_1/\aRH)^{(1-\beta)/\beta}$ besides other less important factors of $\aRH$. Also, using Eq.~\eqref{a1} we obtain the first part of the exponent since $a_1/\aRH = (\TRH/T_1^{\rm std})^{(3+\gamma)/(3+\gamma\alpha)}$. Since $1/\beta>1$ for the NSCs considered here (and indeed for all additional fluids with a non-tachyonic equation of state $w_\phi > -1$), the effect of increasing $H_1$ (reducing the coherence length or patch of the typical minicluster) is stronger than the mere larger density from projecting $\rho_{\rm CDM}$ earlier in time. 

The fact that the quantities $1/\beta-1$ and $H_1$ are the largest among the cosmologies considered explain why in the Kination cosmologies the minicluster mass and radius can attain such low values shown in Fig.~\ref{fig:miniclustermassradius}. 

The small difference between a red-shifting, non-decaying Kination field and a decaying one is due to the slower decrease of the temperature with scale factor, due to the entropy production. Indeed, entropy production is the reason why $a_1^{\rm KD}<a_1^{\rm K}$ (see discussion around \eqref{eq:hierarchy}) and thus $(H_1a_1)^{\rm K} < (H_1a_1)^{\rm KD}$ because $H_1a_1\propto 1/a_1^{1/\beta-1}=1/a_1^{2}$ for both kination types. This effect is captured by the small difference in the exponent in Eq.~\eqref{mcmass} between the two cases, which contains the factor $(3+\gamma)/(3+\gamma \alpha)$ which is $\sim 7/6$ for KD cosmology while it is equal to one for Kination cosmology. 

The most relevant parameter in determining the minicluster mass turns out to be the equation of state of the component dominating the expansion at $T_1$, rather than whether entropy is conserved or not. Since $1/\beta-1>0$ for all the NSCs, this seems to suggest that the minicluster mass increases with the reheating temperature. Indeed, this happens for 
\begin{equation}
	\beta < \frac{1}{2} + \frac{\gamma(1-\alpha)}{6+\gamma(1+\alpha)}, 
\end{equation}
which is satisfied in the case of Kination cosmology ($\beta=1/3$), but not for the MD cosmology ($\beta=2/3$), which is an exception to the main trend. In MD, the value $1/\beta-1=1/2$ is so small that the other term $(-1)$ in the exponent in Eq.~\eqref{mcmass} dominates and changes the trend: the minicluster mass decreases with increasing $\TRH$, although the maximal value of the minicluster mass in MD is only one order of magnitude larger than the standard results.

\subsubsection{Structure formation during the NSC period} \label{Structure formation during the NSC period}

In this subsection we consider whether some gravitationally-driven structure formation could already happen at very early times, during the NSC period. If the axion is the CDM particle, the energy density in the axion field at the onset of oscillations is extremely small compared to the background radiation, typically by a factor $a_{\rm eq}/a_1\sim 10^{10}$, see also Appendix~\ref{A note on primordial black hole formation}. Therefore, the evolution of axion density perturbations is mostly sourced by perturbations of the background fields ($\phi$ and radiation), rather than by axion perturbations themselves.

Let us first briefly discuss the standard case of radiation domination together with the non-standard Kination domination. We will adhere to a simple Newtonian picture valid for sub-horizon perturbations, as super-horizon perturbations change at most by factors of $\mathcal{O}\(1\)$. Let us consider the Fourier modes of the density fluctuations of the background density $\delta_\phi=\rho_\phi/\bar\rho_\phi -1$ (a comoving wavenumber $k$ labelling modes is implicit). Small fluctuations in the linear regime evolve according to 
\be
	\ddot \delta_\phi +2 H \dot \delta_\phi + \(\frac{c_s^2 k^2}{a^2}-\frac{4\pi \bar\rho_\phi}{\Mp^2}\)\delta_\phi = 0 . 
	\label{eq:background_fluctuations}
\ee
where the sound-speed is essentially given by the equation of state ($c_s^2 \approx w_\phi$). Perturbations with $k$ above the Jeans momentum $k_J/a=\(4\pi \bar\rho_\phi/c_s^2\)^{1/2}$ oscillate with damped amplitude while those below it can suffer gravitational amplification leading to collapse. Using the Friedman equation, the latter term in Eq.~\eqref{eq:background_fluctuations} can be written as ${4\pi \bar\rho_\phi}/{\Mp^2}=3H^2/2$. In both the cases of radiation- or Kination-dominated scenarios, the Jeans momentum decreases in time and once a perturbation enters the horizon, $k/a>H$, it begins oscillating with an amplitude $\propto 1/\sqrt{a}$. The situation is completely different in a matter-dominated background. In the ultra-cold limit, for which $w_\phi \to 0$, perturbations start growing as $\delta_\phi\propto a$ as soon as they enter the horizon. The extent up to which our background perturbations can source the evolution of axion perturbations is determined by: 1) their absolute size and 2) the speed of the universe expansion. 

The size of the background perturbations at the small scales of our interest is relatively unconstrained. 
Observations of the CMB anisotropies, Lyman-$\alpha$ forest, large-scale structure and other cosmological probes measure 
the dimensionless power spectrum of curvature perturbations $\Delta^2 \propto |\delta|^2\sim 10^{-9}$ in the range $1 {\rm \, Gpc}^{-1} \lesssim k\lesssim 3 {\rm \,Mpc}^{-1}$~\cite{Nicholson:2009pi,Nicholson:2009zj,Bird:2010mp}. The absence of spectral distortions of the CMB Planck distribution can be used to constrain $\Delta^2 \lesssim 10^{-5}$ up to $k \lesssim {(100 \, \rm pc)}^{-1}$~\cite{Chluba:2012gq,Chluba:2012we,Chluba:2013dna,Chluba:2019kpb}. The effects on the yields of Deuterium and Helium from BBN can also be used to extend this bound even further, $\Delta^2 < 0.007$ for ${(100 \,\rm pc)}^{-1} \lesssim k \lesssim {(10 \,\rm pc)}^{-1}$~\cite{Jeong:2014gna}. The absence of effects of primordial black holes has also been used to constraint $\Delta^2 \lesssim 0.01-0.1$ above that range~\cite{Josan:2009qn}. Additional, model-dependent results~\cite{Bringmann:2011ut, Aslanyan:2015hmi} constrain the dimensionless power spectrum down to $\Delta^2 \lesssim 10^{-7}$ in the range  $1{\rm \, Mpc}^{-1} \lesssim k \lesssim {0.1 \,\rm pc}^{-1}$ if the DM is made of WIMPs which would form ultracompact mini-haloes where annihilations and the subsequent emission of gamma-rays could have been detected. Needless to say, this latter case does not affect the scenario under discussion here, since we do not consider WIMP as the DM candidate in this work.

The growth of axion CDM perturbations at sub-horizon scales sourced by the background perturbations in the Newtonian limit can be described by~\cite{Vernizzi:2003vs}
\be
	\ddot \delta_A +\(2 H +\frac{\dot m}{m}\)\dot \delta_A - \frac{3}{2}H^2\delta_\phi = 0,
	\label{axipert}
\ee
in which the term $\dot m/m$ is required from momentum conservation in the event of a time-varying mass. Since the axion mass depends on temperature due to the instanton effects around the QCD phase transition, see Eq.~\eqref{axionmass}, the term $\dot m/m \approx \alpha \gamma H$ turns out to be extremely important for correctly describing perturbations in the axion field, since it is of the same order as the term $2H$. In particular, the time-dependence of the axion mass plays a relatively significant in the case of the Kination cosmology.

If the background cosmology is described by either the radiation or the Kination scenarios, the density fluctuations given by Eq.~\eqref{eq:background_fluctuations} are oscillating and decaying, so that their role in Eq.~\eqref{axipert} can be neglected. For this, setting $\delta_\phi\simeq 0$ in Eq.~\eqref{axipert} gives a constant solution $\delta_A \sim {\rm const}$, plus a time-dependent solution
\be
	\delta_A \propto
	\begin{cases}
		t^{1-\beta(2+\alpha\gamma)},  &  \beta(2+\alpha\gamma) \neq 1\\
		\log t,   & \beta(2+\alpha\gamma) = 1 . 
	\end{cases}
\ee
In either the radiation-dominated or the Kination-dominated scenarios, the argument of the power-law is smaller than zero so that no growing mode exists. This is a direct implication of the inclusion of the term $\dot m/m$, which is non-zero until the axion mass stops growing around temperature $\TL$, so that its contribution is substantial whenever $\TRH \gtrsim \TL$ in Kination scenarios. Had we omitted the term $\dot m/m$, which is equivalent to setting $\gamma \to 0$, we would have obtained the well-known logarithmic growth of matter perturbations during radiation domination or the not-so well known \emph{linear growth} of CDM perturbations during Kination $\delta_{\rm CDM}\propto t^{1/3}\sim a$, which is valid in WIMP CDM scenarios~\cite{Redmond:2018xty}.

Indeed, the authors of Ref.~\cite{Redmond:2018xty} argue that the linear growth is essentially due to the free-streaming of CDM particles after the kick received by the gravitational field upon horizon crossing. The comoving distance free-streamed by such particles has exactly the same time dependence,
\be
	d= \int_{t_i} v_i\frac{dt}{a}\sim \int_{t_i}\frac{p_i}{m_{ai}}\(\frac{a_i}{a}\)^{\alpha\gamma+1}\frac{dt}{a}\sim t^{1-\beta(2+\alpha\gamma)}
\ee 
where we used that momentum redshifts as $p_i/a$, the axion mass grows as $m_A\propto 1/T^\gamma\propto a^{\alpha\gamma}$ and $a\propto t^{\beta}$. In summary, as long as the axion mass continues to increase, axion CDM perturbations do not grow during radiation-, KD-, or Kination-dominated cosmologies. Once the axion mass reaches a constant value, we recover the standard $\log a$ growth valid during radiation domination\footnote{The fact that axion CDM perturbations do not grow during the Kination stage facilitates the treatment for axion CDM, since the growth of perturbations in Kination cosmology is not understood beyond the linear regime~\cite{Redmond:2018xty}, and gravitational collapse has never been proven.}.

The situation can be entirely different in an early period of matter domination. In the following, we refer to the non-relativistic matter dominating the expansion rate of the universe as Dominating Decaying Matter (DDM). During such a period, the DDM background perturbations grow linearly and undergoes gravitational collapse approximately when becoming non-linear. Solving Eq.~\eqref{axipert} for the axion perturbations when the growing solution of Eq.~\eqref{eq:background_fluctuations} is considered gives
\be
	\delta_A \propto \delta_\phi \propto t^{2/3}.
\ee
Indeed, the situation is beautifully similar to the growth of baryon perturbations after recombination in standard $\Lambda$CDM cosmology. Before recombination, the baryon fluid is tightly coupled to the photon fluid. Because of the large effective sound-speed, all relevant modes are above the Jeans wavenumber and baryons cannot clump. However, (sub-horizon) CDM perturbations grow linearly as soon as the universe becomes matter-dominated. When the temperature drops enough for most protons to form neutral Hydrogen, baryons effectively decouple from the CMB, feeling a sudden drop of pressure, which allows them to fall rapidly into the already deep potential wells dug by CDM. 

Analogously, axion fluctuations cannot efficiently grow at early times because the axion mass is so small that axions are relativistic or frozen due to the Hubble friction. Meanwhile, DDM field perturbations start to grow as soon as they enter the horizon during matter domination. When axions abruptly start to behave as CDM, they would immediately fall into the potential wells of the additional DDM. In the analogy depicted, axions play the role of baryons and DDM that of the standard DM after recombination. Sadly, the analogy stops here: firstly, the axion density field already possesses large fluctuations of its own which are typically not correlated with those of the background field; secondly, DDM is decaying, so it stops being relevant for the overall expansion of the universe at $\aRH$ and it would be completely absent after $t\gg \Gamma_\phi^{-1}$. Axions dragged into DDM over-densities would suddenly find themselves left alone, freed from the gravitational pull of the DDM. If the DDM perturbations were linear, axions just continue their free-streaming inwards of the former DDM perturbations, but if the DDM perturbations had already become non-linear the axions would free-stream away from them.

To understand qualitatively the impact on miniclusters, recall that soon after the onset of oscillations, the spectrum of axion DM fluctuations at small scales\footnote{These are the fluctuations of the axion DM field at small scales due to the misalignment and string emission contribution. In the inflationary cosmology, we expect an uncorrelated additive contribution from the adiabatic temperature fluctuations imprinted in the plasma by the inflaton fluctuations. The temperature fluctuations get imprinted as fluctuations in the axion DM density due to the later's dependence on $T_1$. } can be parametrised as~\cite{Vaquero:2018tib}
\be
	\Delta_{\rm a}^2(k) = \frac{k^3}{2\pi^2}\frac{\langle |\delta_A(k)|^2\rangle}{V}
	\simeq \begin{cases}
	0.03(k/k_1)^3, & \!\!k\lesssim k_1\\
	\sim 1, & \!\! k \in (10 \!-\! 100) k_1\\
	(k_1/k)^\zeta, & \!\! k \gg k_1,
	\end{cases}
\ee
where $V$ is the volume and $\zeta\sim 0.5$ but probably decreasing with $k$. The characteristic comoving wavenumber $k_1$ is defined as the wavenumber entering the horizon at $a_1$, 
\be
	k_1 = H_1 a_1. 
\ee
Although this spectrum was derived from simulations in a radiation-dominated background, we take it as rough representative of the axion CDM initial conditions around $a_1$ also for NSC cosmologies, since numerical results in a matter-dominated scenario are not yet available.

The fact that standard axion miniclusters with overdensities $\Phi \sim O(1)$ form with a typical mass of $\sim \rho_A /(k_1/a_1)^3$ is related to the power spectrum decreasing very sharply at scales $k < k_1$, {\it i.e.} length scales larger than $1/H_1$ at $a_1$. At small length-scales, $k>k_1$, fluctuations are already non-linear and one thus expects the formation of ensuing ``small" miniclusters potentially denser, but the largest physical objects are determined by $k\sim k_1$, which is a consequence of the low-$k$ cut-off $\Delta_A^2 \propto k^3$.

In a MD period, fluctuations in the axion field grow even larger than the estimates in Sec.~\ref{Impact of the horizon size} if the perturbations in the DDM are large enough, {\it i.e.} $\delta_\phi > \delta_A(t_1)$. In particular, in order to change the typical minicluster size, they should be able to shift down the typical wavenumber at which $\Delta_A^2\sim O(1)$. In the following, we will therefore concentrate in modes with wave numbers smaller than $k_1$,
\be
	k < k_1, 
\ee
which are still linear at $a_1$. 
If the DDM perturbations are adiabatic perturbations from a vanilla slow-roll inflaton, we could extrapolate the measured perturbations at CMB scales $\Delta^2\sim 10^{-9} (k/k_{\rm CMB})^{n_s-1}$ with $n_s\simeq 0.968(8)$~\cite{Ade:2015xua}. These fluctuations are so small that they would not affect the typical size of axion miniclusters in NSCs\footnote{They could affect substructure and the size of large clusters of miniclusters, though, but these do not have clear observational consequences yet studied.}. Large adiabatic fluctuations are not entirely unexpected in exotic models of inflation, where they can be originated from features in the inflaton potential~\cite{Starobinsky:1992ts} or inflection points~\cite{Polnarev:2006aa}. They can also originate from particle production~\cite{Barnaby:2009dd}, waterfall transitions~\cite{Linde:1993cn}, and other exotic phenomena, see e.g. Refs.~\cite{Barnaby:2010vf, Chluba:2015bqa}. In practice, the amplitude at very small scales is observationally constrained mostly from the requirement to avoid an overproduction of primordial black holes. 

Assume that DDM has an adiabatic spectrum of amplitude consistent with the PBH limits $\Delta_\phi^2\sim 0.01$. The DDM density fluctuation $\delta_\phi(k)$ entering the horizon at $a_k\in (a_1,\aRH)$ experiences linear growth so that the power spectrum at later times is 
\bea
\label{deltaphi}
	\Delta_\phi^2(a) \sim \Delta_\phi^2(a_k)\(\frac{a}{a_k}\)^2 
	\sim 0.01 \(\frac{a}{a_1}\)^2 \(\frac{k}{k_1}\)^4,
\eea
where we used the relation $H=H_1(a_1/a)^{3/2}$ valid for MD cosmology which leads to $a_k/a_1=(k_1/k)^2$. For axion perturbations that are linear at $a_1$, the largest possible growth happens for those that enter the the horizon
earliest, {\it i.e.} $k\sim k_1$, for which $\Delta_\phi^2$ grows by a factor $(\aRH/a_1)^2$. This quantity is determined by our assumption of having the totality of CDM in axions. In the MD scenario we have 
\be
	\(\frac{\aRH}{a_1}\)^{2} = \(\frac{T^{\rm std}_1}{\TRH}\)^{28/9} \sim 10^8,
\ee
where the numerical factor has been estimated for the choice $\TRH = 5\,$MeV and $\at \sim O(1)$.

In the modified MD scenario described, the mass of the typical miniclusters would be given by the size of the largest DDM fluctuation that 
became $\sim O(1)$, dragging CDM axions with it. Here, we refer to the associated momentum $k'_1$ for its similar role to $k_1$ in setting the value of the minicluster mass, without any relation to Eq.~\eqref{eq:111}. The maximum effect takes place at $a=\aRH$. Solving for $k_1'$ in the expression $\Delta_\phi^2(\aRH,k_1')=1$, \eqref{deltaphi}, we obtain
\bea
	k'_1 &\sim& k_1 \(\frac{a_1}{\aRH}\)^\frac{1}{2}\(\Delta_\phi^2(a_k) \)^{-1/4} \sim \nonumber\\
	&\sim& 3 k_1 \(\frac{\TRH}{T^{\rm std}_1}\)^{28/36} \(\frac{0.01}{\Delta_\phi^2(a_k)} \)^{1/4},  
\eea
which can be up to a factor of $\sim 30$ smaller than $k_1$ for the largest values of $\Delta_\phi^2(a_k)\sim 0.01$ and $T_{1}^{\rm std}\sim 2\,$GeV. Note that $k_1'>\HRH \aRH$ if $\Delta_\phi^2(a_k)$ is small enough so that the black-hole constraint is still respected. In this case, the typical minicluster has a size $R_1'\sim (k'_1/\aRH)^{-1}$ at $\aRH$, and thus a mass, 
\be
	M'_c = \frac{4\pi}{3}(1+\Phi_\phi) \rho_{\rm CDM}\(\frac{1}{k'_1}\)^3 = M_{c,0} \(\frac{k_1}{k'_1}\)^3,  
\ee
where $M_{c,0}$ is the minicluster mass in the NSC if DDM fluctuations are negligible $\delta_\phi\sim 0$. Summing up, miniclusters formed during MD and hosted in DDM perturbations can be a factor up to $\sim 30$ times larger and $\sim 30^3$ heavier due to early structure formation during the NSC. However, this requires large DDM fluctuations (not as large as to produce too many black-holes). A more conservative value $\Delta_\phi^2(a_k)\sim 10^{-5}$ yields a more modest enhancement $M'_c \sim 10^2M_{c,0}$ and only for the most extreme values of the reheat temperature. In this scenario, structures have time to become strongly non-linear above $k'_1$ so these it is likely that these typical miniclusters have a lot of substructures. 

It is tempting to think that the substructures of the large $M'_c$-mass miniclusters are something like the previously discussed $M_c$-mass objects ($M_c$ given by $H_1$). However, we think that this would not be the case. The axion CDM inhomogeneities are so small compared to the background that they play no dynamical role. The driving perturbations are those of the DDM, which correspond to slightly (or not so-slightly) different DDM densities which grow and collapse into DDM halos that presumably merge hierarchically from small to large scales. The axion overdensities that would have later lead to $M_c$ miniclusters just follow the DDM gravitational potential. Since these inhomogeneities have negligible self-gravity, they would probably be disrupted inside DDM halos after a dynamical time scale $\sim H^{-1}$ once DDM becomes non-linear.  

The comoving scales that can be significantly affected by the NSC period are bounded from below by the mode that enters the horizon at the minimum reheating temperature, $\TRH\sim 5\,$MeV,
\be
	k > k_{{\rm min} \{\TRH\}} = \left.\HRH \,\aRH\right|_{\TRH=5 \rm \,MeV} \sim (20 {\rm pc})^{-1} .   
\ee 

\subsubsection{Observational parameters}

\begin{figure}[t!]
	\includegraphics[width=0.5\textwidth]{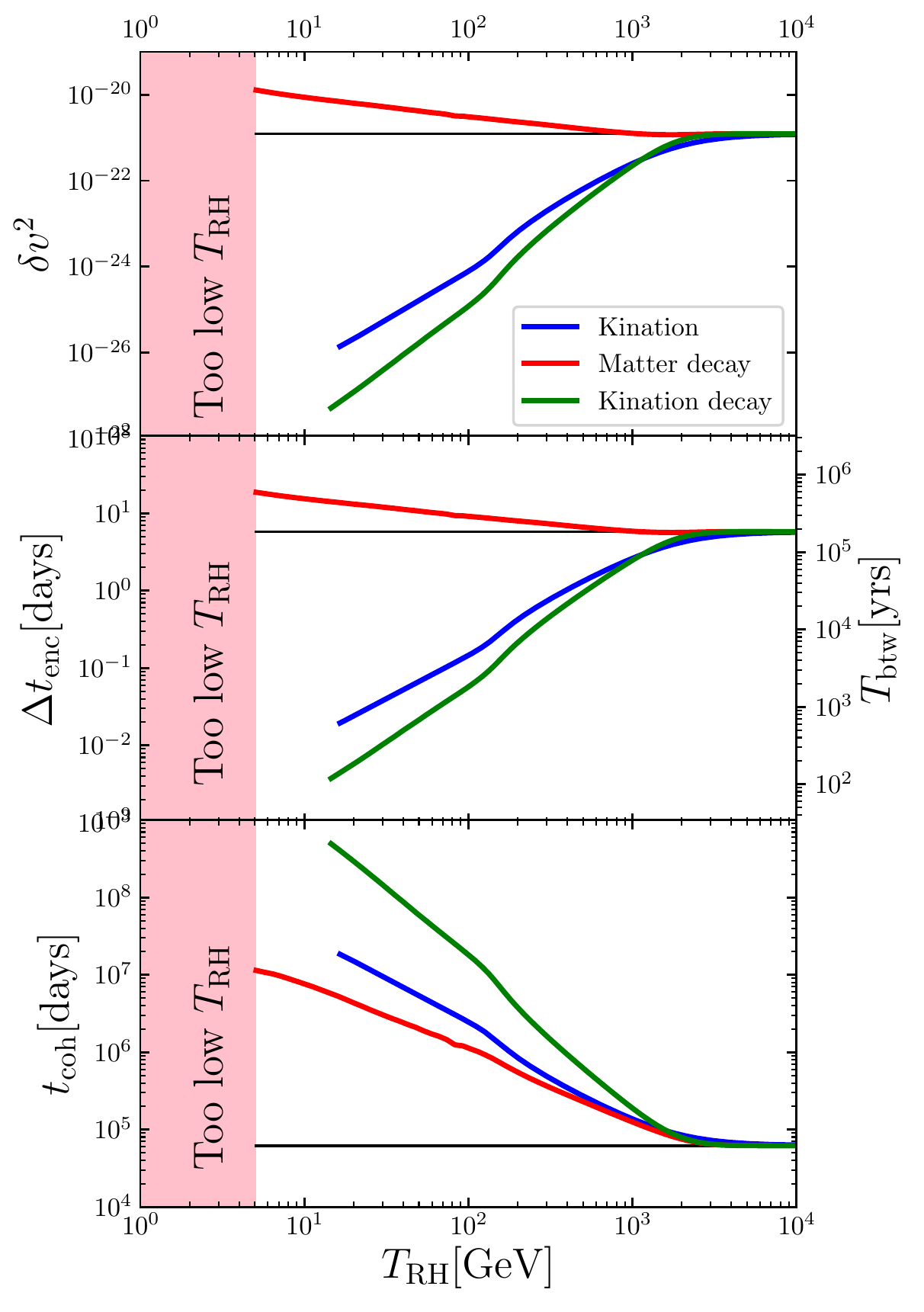}
	\caption{
	Top panel: the velocity dispersion of a minicluster as a function of the reheat temperature for different cosmological models before nucleosynthesis. 
	We have assumed that 100\% of the CDM is in axions with $\at = 10$. 
	Colour coding is the same as in \mbox{Fig.~\ref{fig:miniclustermassradius}}.
	Middle panel: typical duration in days of a single encounter of a minicluster with the Earth. 
	The right vertical axis gives the mean time interval (in years) between two Earth encounters with a minicluster  
	with the same settings as in the Top panel. 
	Bottom panel: coherence time of the axion field in the minicluster. Note that it is always sensibly larger than the  time of an encounter. }
	\label{fig:miniclusterencounter}
\end{figure}

\emph{Velocity dispersion-}
The clustering of axion CDM has observational consequences, since detectors would be triggered by a larger energy density when the Earth passes through one such substructures. In fact, the local energy density of CDM is estimated to be $\rhS \approx 0.4{\rm\,GeV/cm^3}$, see Ref.~\cite{Read:2014qva}, while the density obtained for the minicluster is given in Eq.~\eqref{density_clusters}. The density enhancement during the encounter with a minicluster can then reach 
\be
	\frac{\rho_c}{\rhS} \simeq 1.7\times 10^7 \, (1+\Phi)\Phi^3.
	\label{density_clusters1}
\ee
The velocity dispersion of the axions in the minicluster $\delta v^2$ is directly related to the coherence time of the axion field $t_{\rm coh}$ during such encounter. The latter quantity is a key parameter in an axion search experiment. Both these quantities can be computed from the gravitational potential by the virial theorem
\be
	\delta v^2 \sim \frac{M_c}{\Mp^2 R_c}.
	\label{eq:velocitydispersion_def}
\ee
In the standard cosmological model, once the deviation from the size of the horizon $r = R_1\,H_1 =\mathcal{O}(1)$ is introduced, this gives
\bea
	\delta v^2_{\rm std} &\sim& 2\times 10^{-19} r^2 \Phi(1+\Phi) \(\frac{50 \mu\rm eV}{m_A}\)^{0.3336}, \label{eq:velocitydispersion}\\
	\delta t^{\rm std}_{\rm coh} &\sim& \frac{1}{m_A\, \delta v^2} = 0.9\times10^8{\rm \, s}\frac{1}{r^2\Phi(1+\Phi)}
	\(\frac{50 \mu\rm eV}{m_A}\)^{0.666}, \label{eq:coherencetime}
\eea
where in Eq.~\eqref{eq:coherencetime} we have assumed that the axion mass is equal to $m_{\rm CDM}$. As we discussed in Sec.~\ref{Impact of the horizon size}, the mass and the radius of a minicluster computed in a NSC are modified by the same parameter $1/\(H_1a_1\)$. Thus, the ratio of the velocity dispersions and the ratio of the coherence time when computed in the NSC and in the standard cosmology read
\bea
	\frac{\delta v^2}{\delta v^2_{\rm std}} &\simeq& \(\frac{M_c}{M_c^{\rm std}}\)^{2/3} = \,\left(\frac{\TRH}{T_1^{\rm std}}\right)^{2\(\frac{1-\beta}{\beta}\frac{3+\gamma}{3+\alpha\gamma}-1\)},\\
	\frac{\delta t_{\rm coh}}{\delta t^{\rm std}_{\rm coh}} &\simeq& \left(\frac{\TRH}{T_1^{\rm std}}\right)^{3-\frac{1+\beta}{\beta}\frac{3+\gamma}{3+\gamma\alpha}},
\eea
where the computations for $\delta t_{\rm coh}$ involves using $m_{\rm CDM}$ from Eq.~\eqref{gorgi}. Remarkably, the coherence time in the NSCs we have considered seems to be always longer than in the standard scenario. Referring to Eq.~\eqref{eq:coherencetime}, the axion mass in both the Kination and KD scenarios is larger than in the standard case, however this trend is counterbalanced by the large decrease in the minicluster mass which lowers the velocity dispersion. On the contrary, in the MD scenario the increase in the velocity dispersion due to the more massive miniclusters in this scenario is modest, while the axion mass would be lower by orders of magnitude with respect to the standard value. Our numerical calculations for the velocity dispersion and the coherence time are shown in \mbox{Fig.~\ref{fig:miniclusterencounter}}.

\emph{Encounter rates with the Earth.-}
Assuming the DM mass of the Milky Way as $\MW \sim 10^{12}\,\Ms$, the number of miniclusters in the halo in the standard scenario is, 
\begin{equation}
	N^{\rm std}_c = \frac{M_{\rm MW}}{M_c^{\rm std}} \sim 10^{23} \(\frac{m_A}{50 \mu\rm eV}\)^{0.50},
	\label{eq:standard_number}
\end{equation}
assuming $1/(r^3(1+\Phi))\simeq 1$ on average. Since we take most of the CDM in the form of miniclusters, the local number density of axion miniclusters is then
\begin{equation}
	\nS^{\rm std} \simeq \frac{\rho_{\odot}}{M_c^{\rm std}} \sim 10^9 \(\frac{m_A}{50 \mu\rm eV}\)^{0.50}{\rm \, pc^{-3}}.
	\label{eq:localdensity}
\end{equation}

During a complete revolution around the galactic halo, the Solar System transverses a length $l = 2\pi\,\rS$, where $\rS = 8.3{\rm~kpc}$ is the distance of the Solar System from the galactic centre. On this path, the Earth encounters a number of miniclusters equal to,
\be
	N_{\rm enc}^{\rm std} = (2\pi\,\rS)\,(\pi\,(R_c^{\rm std})^2)\,\nS^{\rm std} \sim 600 
\(\frac{m_A}{50 \mu\rm eV}\)^{0.1668}.
	\label{eq:numberofencounters}
\ee
Given a galactic year $\tS = 2\pi \rS/\vS \sim 230{\rm~My}$, the time between two encounters can be estimated as, 
\begin{equation}
	T_{\rm btw}^\std = \frac{\tS}{N_{\rm enc}^\std}  
	\sim 4\times 10^5 {\rm years} \(\frac{50 \mu\rm eV}{m_A}\)^{0.1668}.
	\label{eq:timebetweenencounter}
\end{equation}

During the encounter, the energy density in the minicluster is enhanced by the factor given in Eq.~\eqref{density_clusters1}.
Given the size of the minicluster in the standard scenario in Eq.~\eqref{size_std} and the velocity of the Solar System around the Galactic centre $\vS \sim 230~$km/s, which coincides with typical virial velocities in the DM halo, the typical encounter has a duration 
\be
	\Delta t_{\rm enc}^{\rm std} = \frac{2R_c^{\rm std}}{\vS} 
	\sim  6{\rm \,days} \frac{r}{\Phi }\(\frac{m_A}{50 \mu\rm eV}\)^{0.1668},
	\label{eq:encountertime}
\ee
where we have assumed that the relative velocity between the Solar System and the minicluster is of the order of $\vS$. According to Eq.~\eqref{eq:coherencetime}, the axion field is coherent during the whole encounter, since $\delta t_{\rm coh} \gg \Delta t_{\rm enc}$. The situation for standard miniclusters is a bit unfortunate. Although the enhancement of the signal  in an axion search experiment during an encounter would be extraordinary, the collision rates are so small that it seems quite hopeless to expect a minicluster encounter. 

Let us now turn our attention to the case of an early NSC. In the Kination and KD cosmologies, the typical minicluster is generally lighter and smaller than in the standard cosmology, $M_c < M_c^\std$ and $R_c < R_c^\std$. According to Eq.~\eqref{eq:standard_number}, we thus expect a larger number of miniclusters in the halo, each with a smaller encounter probability with the Earth. On the contrary, miniclusters formed during a MD cosmology are slightly more massive and larger than what obtained in the standard scenario. We estimate the number of encounters of a minicluster with the Earth $N_{\rm enc}$ and the average lapse time between two encounters in a NSC by using the results in Eqs.~\eqref{eq:localdensity} and~\eqref{eq:numberofencounters} and the expression in Eq.~\eqref{eq:timebetweenencounter}, to obtain
\bea
	\frac{N_{\rm enc}}{N_{\rm enc}^\std} &\simeq& \(\frac{T_{\rm btw}}{T_{\rm btw}^\std}\)^{-1} \propto \(\frac{R_c}{R_c^\std}\)^2\,\frac{M_c^\std}{M_c} = \nonumber\\
	&=& \frac{R_c^\std}{R_c} \propto \left(\frac{\TRH}{T_1^{\rm std}}\right)^{1-\frac{1-\beta}{\beta}\frac{3+\gamma}{3+\alpha\gamma}},
	\label{eq:numberofencounters_NSC}
\eea
where in the last expression we used Eq.~\eqref{mcmass}. Chances to encounter a minicluster halo are then higher for miniclusters forming during a Kination NSC, with a smaller lapse time $T_{\rm btw}$ between two subsequent encounters. The drawback of the scenario lies in the smaller time for which the encounter lasts, which in the standard scenario is given by Eq.~\eqref{eq:encountertime} and for a NSC gives
\be
	\frac{\Delta t_{\rm enc}}{\Delta t_{\rm enc}^\std} \simeq \frac{R_c}{R_c^\std} \propto \left(\frac{\TRH}{T_1^{\rm std}}\right)^{\frac{1-\beta}{\beta}\frac{3+\gamma}{3+\alpha\gamma}-1}.
	\label{eq:encountertime_NSC}
\ee
In \mbox{Table~\ref{table_parameters}} we have summarised the results for the relevant astrophysical quantities of a typical axion minicluster as obtained in this Section for the NSCs we have considered. We have computed the quantities presented in this Section by assuming that all of the axions clump into miniclusters structures and that axions make up the totality of the CDM budget. In the following, we have set for convenience $\TM = \TRH / {\rm \,MeV}$, and we have fixed $\at = 10$ to account for the relative contribution to the axion energy density from the decay of topological defects. We have also set for simplicity $r = 1$, and we have considered miniclusters that form with an overdensity $\Phi = 1$. For these reasons, the results only depend on the reheat temperature $\TRH$. The mass of the axion for which we have 100\% CDM, here $\mCDM$, differs by various orders of magnitude among the different cosmologies, as first noted in Ref.~\cite{Visinelli:2009kt}. Likewise, the mass and size for a minicluster in either the standard and MD scenarios can have similar ranges.

\bgroup
\def\arraystretch{1.5}
\begin{center}
\begin{table*}[tb]
\centering
\begin{tabular}{|l|c|c|c|c|}
\hline
Scenario \hspace{1cm} & STD \hspace{2cm}  & MD \hspace{2cm} & Kination (no decay) & Kination (w/ decay)\\ 
\hline
$\alpha$ & 1 & 3/8 &1 & 3/4\\
$\beta$ & 1/2 & 2/3 & 1/3 & 1/3\\
$\mCDM~({\rm \mu eV})$ & 100 & $10^{-1}\,\TM^{4/3}$ & $10^4\,\TM^{-1}$ & $10^4\,\TM^{-1}$\\
$T_1~({\rm MeV})$ & $1500$ & $70\,\TM^{5/12}$ & $1500$ &  $600\,\TM^{1/8}$\\
$M_c~(\Ms)$ & $10^{-11}$ & $10^{-9}\,\TM^{-2/3}$ & $10^{-21}\,\TM^3$ & $10^{-24}\,\TM^4$\\
$R_c~({\rm AU})$ & $0.5$ & $\TM^{-2/9}$ & $10^{-4}\,\TM$ & $3\times 10^{-5}\,\TM^{4/3}$\\
Enhancement & $10^7$ & $10^7$ & $10^7$ & $10^7$\\
$\delta t_{\rm coh}~({\rm s})$ & $3\times 10^7$ & $10^{10} \TM^{-8/9}$ & $10^{11}\TM^{-1}$ & $10^{13}\TM^{-5/3}$\\
$\Delta t_{\rm enc}~({\rm days})$ & $5.5$ & $28\,\TM^{-2/9}$ & $4\times 10^{-3}\,\TM$ & $3\times 10^{-4}\,\TM^{4/3}$\\
$N_{\rm enc}$ & $500$ & $10^2\,\TM^{2/9}$ & $10^6\,\TM^{-1}$ & $10^7\,\TM^{-4/3}$ \\
$T_{\rm btw}$ (yr) & $4\times10^5$ & $10^6\TM^{-2/9}$ & $300\TM$ & $20\TM^{4/3}$\\
\hline
\end{tabular}
\caption{The parameters $\alpha$ and $\beta$ describing the various pre-BBN cosmologies: Standard radiation-dominated (STD), matter-dominated low-reheat temperature (MD) cosmology, Kination respectively without or with (KD) the decay of the $\phi$ field considered. For each cosmology, we provide the value of the relevant quantities describing the structure of the axion minicluster and the details of encounter with the Earth, setting $\at = 10$, $\Phi = 1$, and $r=1$. We have defined $\TM = \TRH / {\rm \,MeV}$. $\mCDM$ is the value of the axion mass for which the axion is the CDM particle in the specific cosmology considered, in which the axion field begin the coherent oscillations at temperature $T_1$. $R_c$ and $M_c$ are respectively the radius and the mass of the minicluster. The local CDM energy density is enhanced by the quantity under ``Enhancement'' by the presence of the minicluster. The encounter of the Earth with an axion minicluster would last $\Delta t_{\rm enc}$ days, with a period between two encounters given by $T_{\rm btw}$. $N_{\rm enc}$ is the number of miniclusters encountered by one galactic revolution.}
\label{table_parameters}
\end{table*}
\end{center}
\egroup

\emph{Minicluster streams.-} \label{Minicluster streams}

Tidal axionic streams~\cite{Tinyakov:2015cgg, Dokuchaev:2017} form by the encounter and the subsequent disruption of a minicluster with nearby stars in the disc or by the gravitational field in the halo, similarly to other types of clumps that might have formed in the early universe~\cite{Berezinsky:2014wya}. These structures might be crucial for the direct detection of axions in that the energy density, while diminished with respect to what is attained in a minicluster produced in the standard cosmological picture, can nevertheless be larger than the average DM density $\rho_{\odot}$.

In the standard picture, axion miniclusters are too dense to be disrupted by the gravitational field of the halo, and the main process for their disruption is by tidal stripping after an encounter with a star. Defining a critical impact parameter $b_c$ as~\cite{Goerdt:2006hp, Schneider:2010jr}
\be
	b_c \equiv \sqrt{\frac{M_s R_c}{\Mp^2\,v_{\rm rel} v_c}},
	\label{eq:critical_parameter}
\ee
any minicluster whose impact parameter with a nearby star is $b < b_c$ would be disrupted after a single encounter. In Eq.~\eqref{eq:critical_parameter}, we have defined the velocity dispersion of the minicluster $v_c = \sqrt{\delta v^2}$ from Eq.~\eqref{eq:velocitydispersion_def} and $v_{\rm rel}$ is the velocity of the minicluster relative to the star of mass $M_s$. Given a column density of stars $\Sigma_{\perp} = dM_s/d\mathcal{A}$ for a given area $\mathcal{A}$ in the direction orthogonal to the disc, the total probability of encounter over the period $\Delta t_{\rm obs}$ is then
\be
	p_{\rm disr}(\Phi) = \frac{2 \pi R_c (4\Sigma_{\perp})\Delta t_{\rm obs}}{\Mp^2\,v_{\rm rel} v_c} = \frac{8 \pi \Sigma_{\perp}\Delta t_{\rm obs}}{\Mp\,v_{\rm rel}}\sqrt{\frac{4\pi}{3\rho_c}},
	\label{eq:disruption_probability}
\ee
where the additional factor of four in $(4\Sigma_{\perp})$ has been estimated in Ref.~\cite{Tinyakov:2015cgg} from the averaging over all directions, and the extra factor of two comes from summing up all non-disruptive encounters at impact radii $b > b_c$~\cite{Goerdt:2006hp}. For obtaining the result, we have used the relation between the mass and the radius of a minicluster as in \mbox{Eq.~\eqref{rhoc}}. Notice that the disruption probability depends on the density of the minicluster but not on $M_c$ or $R_c$ separately, so $p_{\rm disr}$ is the same for any cosmological model used. This because in larger miniclusters, the escape velocity of bound axions is also larger by the same factor when the density of the object is constant. Using the expression in \mbox{Eq.~\eqref{rhoc}} for the density of a minicluster and the estimate for the column density $\Sigma_{\perp} \approx \(35\pm 5\)\,M_{\odot}{\rm \,pc^{-2}}$~\cite{Kuijken:1989hu} gives $p_{\rm disr} = 0.1\% (1+\Phi)^{-1/2}\Phi^{-3/2}$ per galactic year.

Owing to the disruption of the minicluster, an enhanced dark matter density $\rho_{\rm stream}(\Phi, t)$ with respect to the background value $\bar \rho_A$ is expected. Axions that have been stripped away from the bound orbits of the minicluster stream away, traveling a distance $L = v_c t$ in a time $t$ and with a density of the stream that drops linearly with time as~\cite{Tinyakov:2015cgg}
\be
	\rho_{\rm stream}(\Phi, t) \!=\! \rho_c \frac{R_c}{v_c t} \!\approx\! 10^7{\rm \frac{GeV}{cm^{3}}}\,(1+\Phi)^{1/2}\Phi^{3/2}\frac{\tau_G}{t_G},
\ee
which again is independent on the cosmological model. We have introduced the characteristic time $\tau_G = 35500\,$yr, while $t_G \sim 12\,$Gyr is the age of the galactic disc. The enhancement $E$ of the axion energy density due to tidal stream then ranges from a minimum $E_{\rm m}(\Phi) = \max\(E,\rho_{\rm stream}(\Phi, t)/\bar\rho_A\)$ to a maximum value $E_{\rm M} = \rho_c/\bar\rho_A$. Following closely Eq.~(4.5) in Ref.~\cite{Tinyakov:2015cgg}, we obtain the rate of stream encounters
\be
	N(E) = \int_0^{+\infty} d\Phi \int_{E_{\rm m}(\Phi)}^{E_{\rm M}} \frac{dE'}{(E')^3}\,\frac{p_{\rm disr}(\Phi)\,f(\Phi)}{2R_c/v_{\rm rel}}\frac{\rho_{\rm stream}(\Phi, t)}{\bar\rho_A}.
	\label{eq:def_NA}
\ee
Here, we have introduced the mass fraction in miniclusters, first estimated in Ref.~\cite{Kolb:1995bu}, although we note that the refined calculations of Ref.~\cite{Vaquero:2018tib} seem to point to a much smaller density of ultra-dense miniclusters. 

\begin{figure}[t!]
\centering
\includegraphics[width=0.5\textwidth]{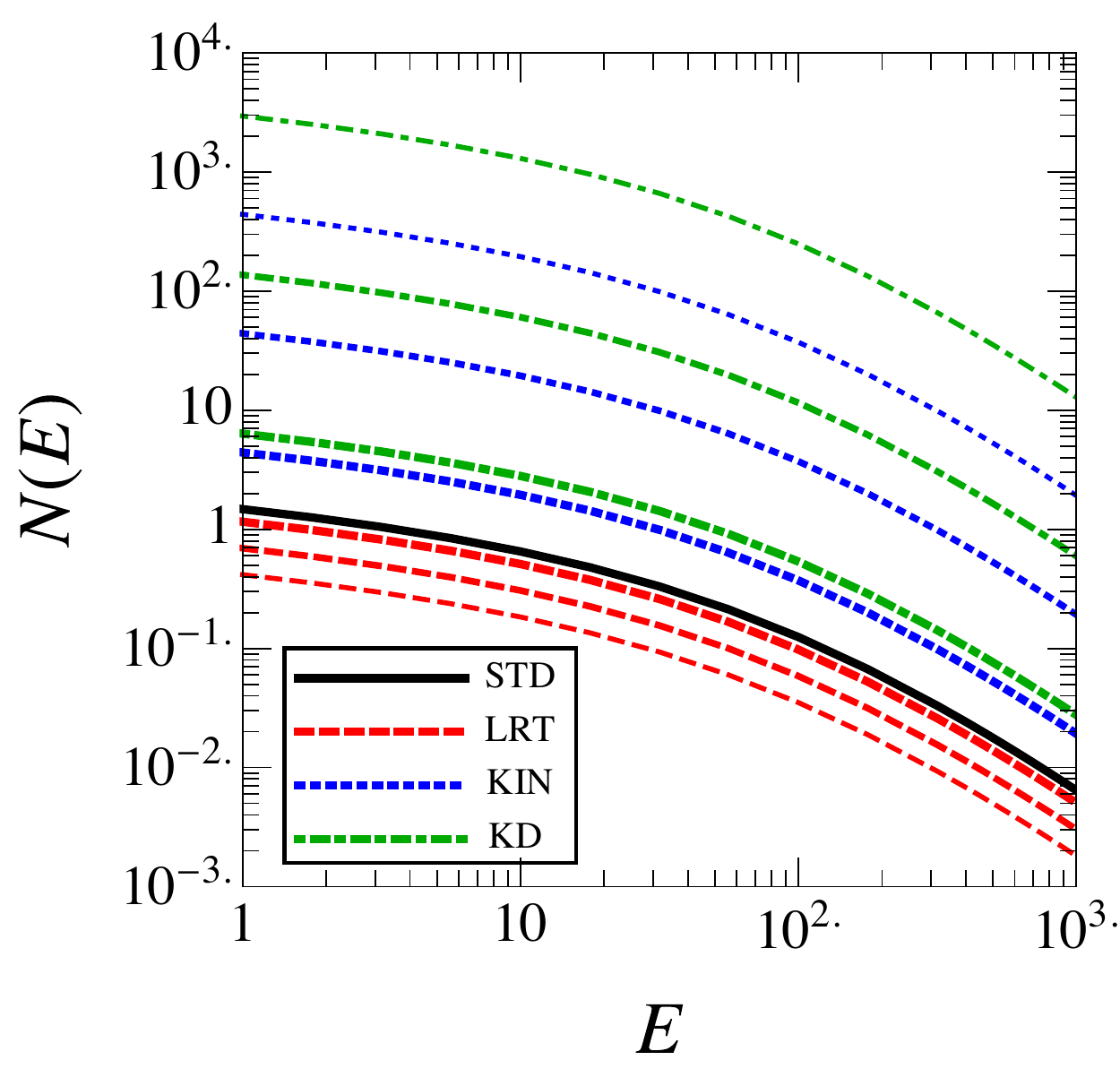}
	\caption{The number of encounters $N(E)$ with an axion stream produced from the tidal stripping of a minicluster, leading to an enhancement larger than $E = \rho/\bar\rho_A$. We have assumed an observation interval $\Delta t_{\rm obs} = 20\,$years and a contribution from topological defects so that $\at = 10$. The black solid line represents the result obtained in the standard cosmological scenario, while the other NSC scenarios are plotted with different colours and dashing: MD (red dashed line), Kination (blue dotted line) and KD (green dot-dashed line). Thicker lines correspond to larger reheat temperatures, corresponding to $\TRH = \(5,\, 50,\, 500\)\,$MeV.}
	\label{fig:plotstream}
\end{figure}

In \mbox{Fig.~\ref{fig:plotstream}}, we show the result obtained from computing the number of encounters $N(E)$ in Eq.~\eqref{eq:def_NA} that yield to an enhancement in density larger than $E$, plotted as a function of $E$. We have fixed the observation period $\Delta t_{\rm obs} = 20\,$years and we have assumed that strings contribute to the total axion energy density so that $\at = 10$. We show the results obtained in the standard cosmological history with a black line, and we have included the contributions from other NSC with different colour codings: MD (red dashed line), Kination (blue dotted line) and KD (green dot-dashed line). Thicker lines correspond to larger reheat temperatures in $\TRH = \(5,\, 50,\, 500\)\,$MeV. The result in the standard cosmology is smaller than what obtained in Ref.~\cite{Tinyakov:2015cgg} because of the more massive miniclusters we used in our model, $M_c^{\rm std} \sim 10^{-11}\,M_{\odot}$, consistently with the analysis in the previous sections. For the same density, the miniclusters we obtain in our analysis are thus slightly larger than what used in previous analyses, and it takes more time to transverse them affecting the denominator in Ref.~\eqref{eq:def_NA}. This reasoning also explains the trend observed in Fig.~\ref{fig:plotstream} for NSC scenarios, for which a more massive and larger minicluster is realised in the MD cosmology while smaller miniclusters are obtained in Kination and KD cosmologies.

\section{Discussion and conclusions} \label{Discussion and conclusions}

In this paper, we have discussed the properties of axion miniclusters emerging in different cosmological scenarios before Big Bang Nucleosynthesis (BBN) took place. In particular, we have considered different scenarios in which the cosmology before BBN was governed by either i) a matter component, ii) a fast-rolling field $\phi$ leading to a Kination period, or iii) a decaying Kination field $\phi$. Using assumptions commonly made in the literature, we have obtained the mass and size of the minicluster, as well as the enhancement in axion density over the local CDM background, in different cosmological setups.

We have sketched the results for the relative quantities describing miniclusters in more detail in \mbox{Fig.~\ref{fig:miniclustermassradius}} as a function of the temperature $\TRH$ at which the modified cosmology transitions to the standard radiation-dominated scenario. In the Figure, we show the mass of the minicluster in the case whether the early cosmological scenario is standard (black lines), matter-dominated (red lines), Kination without (blue lines) and with the decay of the $\phi$ field (green lines). Solid and dashed lines assume $\at=10$ or $\at = 1$, respectively. In order to produce the figure, we fix the relic abundance to the present CDM abundance, so that for each value of $\TRH$ the axion mass is given by Eq.~\eqref{mmm}. We have cut the plots at the value of $\TRH$ for which the axion mass exceeds the bound from the astrophysical considerations or the minimum reheating temperature. In \mbox{Fig.~\ref{fig:miniclustermassradius}}, the right vertical axis gives the size of the minicluster, obtained using the fact that the minicluster density is constant, see \mbox{Eq.~\eqref{size_minicluster}}. For $\TRH \leq T_1^{\rm std}$, the mass and the size are steadily smaller than the standard value for Kination cosmologies, while it is higher than what obtained in the standard scenario for the matter-dominated model. In more details, miniclusters  in the MD cosmology can have a mass is up to two orders of magnitude larger than standard (radius up to $\sim 5$ larger), while in the Kination and KD scenarios the mass can be up to a factor $10^9$ smaller than standard (with a radius up to $10^3$ times smaller). The miniclusters obtained when considering the Kination or the KD cosmologies are lighter and more compact, thus making it more frequent for the Earth to come into the vicinity of these objects. As we obtained in \mbox{Fig.~\ref{fig:massCDM}}, when $\TRH \geq T_1^{\rm std}$ the axion field starts to oscillate in the standard scenario and we recover the standard results.

When we assume that all of the DM is in the form of axions, the typical minicluster density is set by the DM density at matter-radiation equality, $M_c/R_c^3\sim \rho_{\rm eq}$, and does not depend on the early cosmology within our simplified picture. The minicluster mass and radius however can be very different from standard cosmology as they are set by the size of the horizon when the axion field begins to oscillate and becomes non-relativistic. The astrophysical quantities of relevance for  detection tend to depend on different combinations of $M_c$ and $R_c$ and can be very different from the standard scenario: the velocity dispersion $\delta v \propto \sqrt{M_c/R_c}$, the time between encounters with the Earth $T_{\rm btw} \propto M_c/R_c^2$ and the duration of an encounter $\Delta t_{\rm enc} \propto R_c$ are different in non-standard cosmologies for different values of the reheating temperature. In \mbox{Fig.~\ref{fig:miniclusterencounter}} (Top panel) we show the typical duration of a minicluster encounter with the Earth (left vertical axis) and the expected time interval between two consecutive encounters (right vertical axis) as a function of $\TRH$, as well as the velocity dispersion squared (Bottom panel) for the cosmological model studied. For $\TRH \leq T_1^{\rm std}$, both modified cosmologies show detection advantages and disadvantages compared to the standard result. If the axion starts oscillating in a Kination model, the encounter would only last up to a few minutes owing to the small size of the minicluster itself; on the other hand, the frequency of encounter in the Kination cosmology can be enhanced by an $O(10^3)$ factor with respect to the standard case, with the encounters possibly being as frequent as one per a few years. On the contrary, for an axion field that begins to oscillate in a matter-dominated scenario, the encounter would last up to $\approx 50$ days, although one such encounter during a Galactic year would be much more rare. For an axion minicluster forming in the standard cosmology, the velocity dispersion is small enough so that the coherence time of the axion field is much longer than the duration of a minicluster encounter with the Earth. In any modified cosmology we study, the coherence time modifies but not as much as to invalidate the previous statement.

\begin{figure*}[h!]
\includegraphics[width=\textwidth]{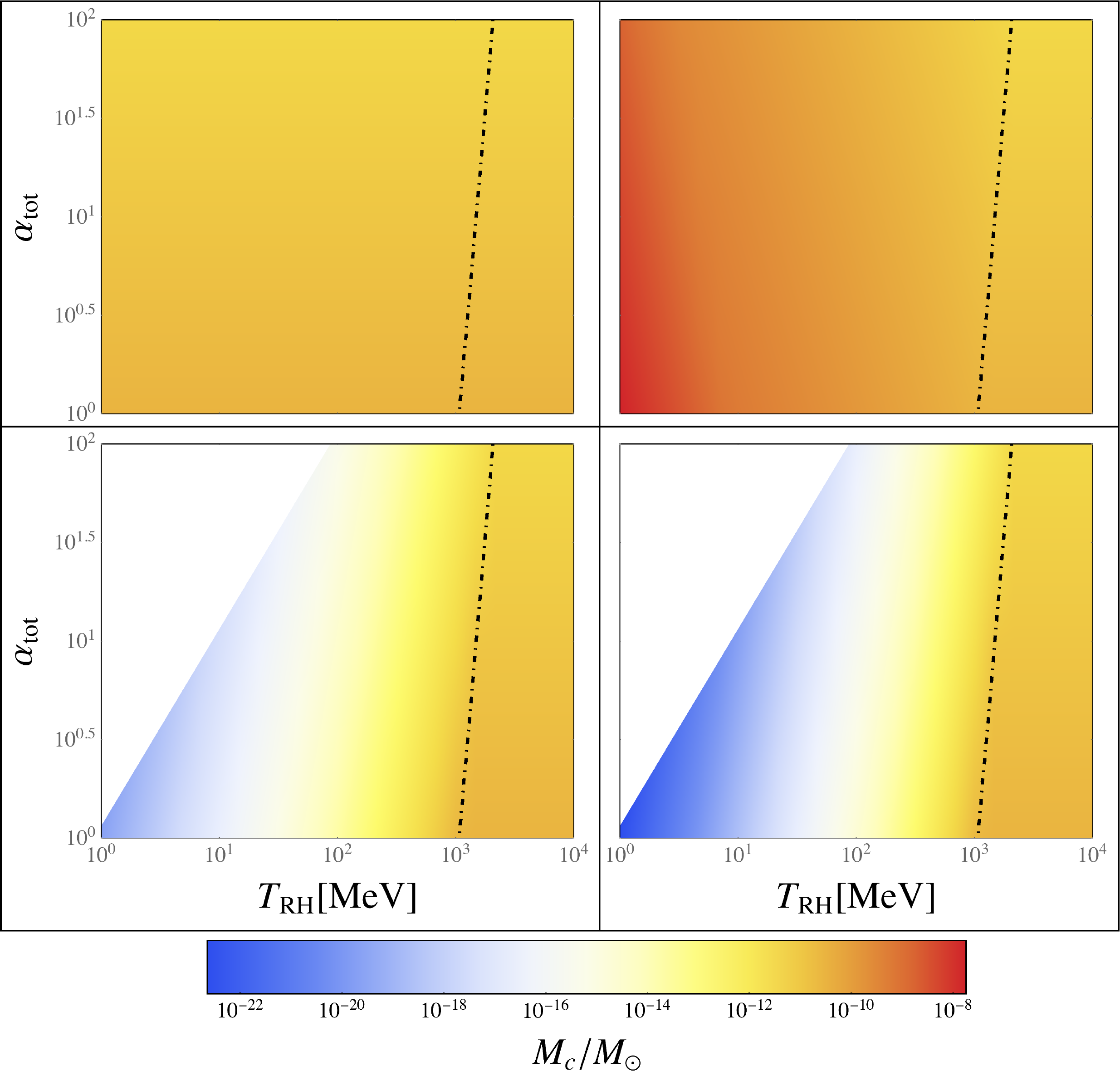}
	\caption{Density plot showing the mass of an axion minicluster, depending on the values of the reheat temperature and the parameter $\at$, for different cosmological models before nucleosynthesis. Top left: Standard scenario. Top right: Low-reheat temperature scenario. Bottom left: Kination scenario. Bottom right: Kination scenario with a decaying $\phi$ field. The dot-dashed line marks the region where the axion field starts to oscillate in the standard scenario (right side) or in the modified scenario (left side).}
	\label{fig:miniclustermass}
\end{figure*}

We discuss the dependence of the solution on $\at$ spanning through various orders of magnitude, since at presence the effective value of this quantity is uncertain. In \mbox{Fig.~\ref{fig:miniclustermass}} we report the density plot showing the mass of the axion minicluster, in units of $\MS$, depending on both $\TRH$ and $\at$. Again, the largest variations in mass are shown for the Kination models, for which the mass of the minicluster ranges between $10^{-22}$ to $10^{-8}$ solar masses over the allowed range. The range over which the mass of the minicluster varies is much more contained in the standard cosmology, for which $M_c \sim 10^{-11}\,\MS$, and in the MD cosmology for which $M_c$ varies by just two orders of magnitude around the standard value. The white region marks the area where the axion mass is excluded by astrophysical considerations. The dot-dashed line marks the region where $\TRH < T_1^{\rm std}$, where the modified cosmology takes place to the left of the dot-dashed line, and the region $\TRH > T_1^{\rm std}$ where the axion field starts to oscillate in the standard radiation-dominated cosmology, for which $M_c$ is given by the value in the standard cosmological scenario. Overall, the actual value of $\at$ does not change much the general picture.

One might question whether an early matter stage, as the one the universe experiences in the MD cosmology, leads to a pre-BBN growth of the structures since adiabatic perturbations in the modulus field entering horizon in such a modified cosmology would grow linearly with the scale factor. This has been considered for CDM seeds in MD models in Refs.~\cite{Erickcek:2011us, Erickcek:2015bda}, and in Kination models in Refs.~\cite{Redmond:2017tja, Redmond:2018xty}. Perturbations in the axion dark matter fluid would be dragged into such primordial perturbations and possibly lead to an early growth. However, such model would be depending on the initial power spectrum of the perturbations in the new field. We have partly addressed this issue in Sec.~\ref{Structure formation during the NSC period} where we have shown that perturbations in the energy density of the axion grow linearly with time only in a MD cosmology, while in Kination the growth is suppressed. This behaviour is peculiar to axion CDM and differs from the WIMP scenario studied in Refs.~\cite{Redmond:2017tja, Redmond:2018xty}. We have been able to estimate the largest mass that could grow into a minicluster in the MD model due to the large fluctuations in the massive scalar field governing the expansion rate in the NSC $\delta_\phi$, by demanding that fluctuations are not so large as to produce an excessive number of primordial black holes. We have found that the radius of the axion minicluster could be enhanced by a factor up to $\sim 30$, corresponding to an enhancement $\sim 30^3 \approx \mathcal{O}\(10^4\)$ in mass with respect to the case in which fluctuations are suppressed $\delta_\phi \simeq 0$. In this scenario, miniclusters would then attain a mass $M_c \sim 10^{-6}\,M_{\odot}$ which is of the same order of magnitude as the mass of the first halos that form in WIMP models from collisional damping and free-streaming that erase density perturbations within that scale\footnote{Depending on WIMP-lepton scattering cross-section, the value of the free-streaming length can also vary by orders of magnitude. The WIMP free-streaming length in a NSC has been estimated in Ref.~\cite{Visinelli:2015eka}.}. We leave further details on the important and interesting issue of an early growth of perturbations to a subsequent work

A further aspect which is worth discussing is the eventual survival of axion minicluster from tidal stripping. Here, we have dealt with these computations in Sec.~\ref{Minicluster streams} following closely the treatment in Ref.~\cite{Tinyakov:2015cgg} and the paper therein in which the issue has been addressed for miniclusters in the standard cosmology. As for any dark matter micro-halo~\cite{Goerdt:2006hp, Schneider:2010jr}, the disruption probability after one passage of an axion minicluster through the Galactic disc is given in \mbox{Eq.~\eqref{eq:disruption_probability}}, which we have shown to be independent on the details of the NSC at the lowest order of the approximation. This result holds because the disruption probability is approximately $p_{\rm disr} \propto \sqrt{1/\rho_c} \approx {\rm const}$. In the simplest model we have discussed, the probability of disruption is then independent on the details of the cosmology and on the details of the physics of the axion. The result $p_s \ll 1$, valid in the standard scenario, is then expected to hold also in modified cosmological histories. We then expect a sizeable fraction of the dark matter axions to be bound into miniclusters even in modified cosmologies, since tidal stripping does not seem to provide a mechanism of disruption of these sub-structures. In any case, we have computed the contribution from the tidal stripping ox axion miniclusters to the local energy density in the form of axion streams, by extending the results discussed in Ref.~\cite{Tinyakov:2015cgg} to a NSC scenario. As we show in \mbox{Fig.~\ref{fig:plotstream}}, the number of encounters $N(A)$ with an axion stream for a given enhancement $A$ is expected to be sensibly larger than in the standard scenario for the early Kination and KD cosmologies. In particular, a number of encounters of the order of $N \sim 10^2$ to $10^3$ are expected even for an enhancement $A \sim 10$ of the local axion density.

For these reasons, we believe it is worth readapting the existing experimental strategies of detecting axion DM to take into account this broad range of minicluster masses and radii shown in \mbox{Fig.~\ref{fig:miniclustermassradius}}. In the event of a discovery, the minicluster size distribution could be a window to the cosmology in the still unexplored era prior to big-bang-nucleosynthesis.

{\bf Note added:} During the completion of the present work, Ref.~\cite{Nelson:2018via} appeared, with their results for an early matter-dominated epoch overlapping with our work.

\begin{acknowledgments}
The authors would like to thank Pierre Sikivie, Sebastian Baum, and Sunny Vagnozzi for the useful discussions and comments that led to the present work. L.V. would like to thank the University of Zaragoza, where part of this work was conducted, for hospitality. L.V. acknowledges support by the Vetenskapsr\r{a}det (Swedish Research Council) through contract No. 638-2013-8993, the Oskar Klein Centre for Cosmoparticle Physics, and support from the NWO Physics Vrij Programme ``The Hidden Universe of Weakly Interacting Particles'' with project number 680.92.18.03 (NWO Vrije Programma), which is (partly) financed by the Dutch Research Council (NWO). J.R. is supported by the Ramon y Cajal Fellowship 2012-10597, the grant FPA2015-65745-P (MINECO/FEDER), the EU through the ITN  ``Elusives'' H2020-MSCA-ITN-2015/674896 and the Deutsche Forschungsgemeinschaft under grant SFB-1258 as a Mercator Fellow.
\end{acknowledgments}

\appendix

\section{Axion Production mechanisms} \label{Production mechanisms}

In this section, we review the axion physics and cosmological production. For an excellent introduction to the subject we refer to Ref.~\cite{kolb:1994early}, while thorough reviews are found in Refs.~\cite{Raffelt:2006rj, Peccei:2006as, Sikivie:2006ni, RevModPhys.82.557, Duffy:2009ig, Ringwald:2012hr, Kawasaki:2013ae, Marsh:2015xka} and in the appendix to Ref.~\cite{Dias:2014osa}.

Populations of cosmological axions are produced through five main mechanisms: thermalisation~\cite{PhysRevLett.59.2489}, the decay of a parent particle~\cite{Higaki:2012ar, Cicoli:2012sz, Cicoli:2012aq, Hasenkamp:2012ii, Conlon:2013isa}, vacuum realignment~\cite{Preskill:1982cy, Abbott:1982af, Dine:1982ah}, the decay of topological string defects~\cite{PhysRevD.32.3172, Battye:1993jv, Harari:1987ht, Hagmann:1990mj, Chang:1998tb, Yamaguchi:1998gx, Hiramatsu:2010yu}, and wall decay~\cite{Nagasawa:1994qu, Chang:1998tb, Hiramatsu:2010yn, Hiramatsu:2012gg, Kawasaki:2014sqa}. Of these mechanisms, only the latter three contribute to a sizeable cold dark matter population. We briefly revise these production methods.

\begin{itemize}
\item Thermal axions

Thermal axions are produced in the early universe mainly through the process $\pi + \pi \to \pi + a$~\cite{Chang:1993gm}. Similarly to neutrinos, thermal axions would contribute the hot dark matter component. For this reason, an upper bound $m_A \lesssim 1{\rm~eV}$ can be placed from the requirement that thermal axions do not overclose the universe~\cite{Hannestad:2003ye, Hannestad:2010yi, Archidiacono:2013cha, DiValentino:2015wba}.

\item Decay of a parent particle

A decaying massive particle or a modulus coupled to the axion field would lead to an increment of the hot dark matter or dark radiation components, through the decay of the modulus into two axions. An effective model for the massive modulus would be a low-energy manifestation of a larger theory involving both supersymmetry and extra dimensions~\cite{ArkaniHamed:1998nn, Randall:1999vf, Binetruy:1999ut, Chung:1999xg, Caldwell:2001ja, Visinelli:2014xsa, Visinelli:2017bny} like a string theory axion~\cite{Svrcek:2006yi, Arvanitaki:2009fg, Acharya:2010zx, Dine:2010cr, Ringwald:2012cu, Cicoli:2012sz, Kim:2013fga, Bachlechner:2014gfa, Halverson:2017deq, Stott:2017hvl, Visinelli:2018utg}. For this reason, dark radiation from a string model~\cite{Cicoli:2012aq, Cicoli:2012sz} is able to constrain string and M-theory compactification scenarios through the change in the effective number of relativistic degrees of freedom $N_{\rm eff}$~\cite{Mangano:2001iu}, with constraints coming from both the CMB polarisation and big bang nucleosynthesis. In some models, the parent particle is a modulus field which, if it dominates the universe, must decay prior Big Bang Nucleosynthesis (BBN) at a reheat temperature~\cite{Kawasaki:1999na, Kawasaki:2000en, Hannestad:2004px, Ichikawa:2005vw, DeBernardis:2008zz}
\be
\label{reheatingconstraint}
	\TRH \gtrsim 5{\rm \,MeV},
\ee
in order to avoid the so-called ``moduli decay problem''~\cite{Coughlan:1983ci, Acharya:2008bk}. The limit on $\TRH$ results from general considerations on the successes of BBN, and it is then a general lower bound below which the universe has to be dominantly filled with radiation. Here we do not treat further the possibility that axions are produced from the decay of parent particles, since axions as dark radiation do not pile up to the present CDM budget.
 
\item Vacuum realignment

Vacuum realignment, one of the main mechanisms to produce a cold axion population, occurs after the breaking of the PQ symmetry that sets the axion field at the bottom of a ``Mexican hat'' potential~\cite{Preskill:1982cy, Abbott:1982af, Dine:1982ah}. Axions are massless from the breaking of the PQ symmetry down to temperatures of the order of the QCD phase transition, when instanton effects generate an effective axion potential~\cite{DiVecchia:1980yfw, diCortona:2015ldu},
\begin{equation}
	V(\theta) = \frac{\Lambda_A^4}{c_z} \(1-\sqrt{1-4c_z\sin^2(\theta/2)}\),
	\label{Vqcd}
\end{equation}
where $\theta = N a/f_A$ is an angular variable, $\Lambda_A^4 = (75.5 \rm \,MeV)^4$ is the topological susceptibility, and $c_z = z/(1+z)^2 = 0.22$ with the ratio of the up and down quark masses $z=m_u/m_d = 0.48$. 
The square of the axion mass at zero temperature is then~\cite{Weinberg:1977ma, Wilczek:1977pj}
\be
	m_A^2 \!\equiv\! \frac{1}{f_A^2}\frac{d^2 V}{d \theta^2}\bigg|_{\theta=0}\!\!\!\!\! = \frac{\Lambda_A^4}{f_A^2}.
	\label{eq:m0}
\ee
We discuss the vacuum realignment mechanism in a generic cosmological scenario in Sec.~\ref{sec_vrm}.

\item Decay of topological strings

Topological strings are produced because the angular variable $\theta$ takes different values at each spatial point after the breaking of the PQ symmetry, through the Kibble mechanism~\cite{Kibble:1976}. After production, the energy density in strings scales with the string unit length, and the string continuously emit low-frequency modes axions which eventually contribute to the present cold dark matter energy density. The actual emission spectrum is crucial in determining the present abundance of cold axions, which is computed in Refs.~\cite{PhysRevD.32.3172, Battye:1993jv} by using an energy spectrum with a sharp peak at the horizon scale, and in Refs.~\cite{Harari:1987ht, Hagmann:1990mj, Chang:1998tb} by using a spectrum proportional to the inverse of the axion momentum $1/q$. Results are often expressed in terms of the ratio $\as = \rst/\rmt$ of the present energy density of cold axions from axionic strings $\rst$ and that from axions produced via the misalignment mechanism $\rmt$. Refs.~\cite{PhysRevD.32.3172, Battye:1993jv} report $\as \sim 200$, while Refs.~\cite{Harari:1987ht, Hagmann:1990mj, Chang:1998tb} report $\as \sim 0.1$, thus the estimation of the CDM axion mass differs by order of magnitudes in the two models. The controversy between these different models is solved with lattice QCD numerical simulations~\cite{Yamaguchi:1998gx, Hiramatsu:2010yu}, which show that the energy spectrum peaks at the horizon scale and is exponentially suppressed at higher momenta. This method yields an intermediate value $\as \sim 10$. However the recent numerical simulations in Refs.~\cite{Klaer:2017qhr, Klaer:2017ond} find an order of magnitude discrepancy with the results in Refs.~\cite{Yamaguchi:1998gx, Hiramatsu:2010yu}, showing that a consensus on the detail on the axion string radiation into a spectrum of axions has not been reached yet. All of the results discussed are valid in a radiation-dominated cosmology, however the value of $\as$ also depends on the properties of the cosmological model before BBN~\cite{Visinelli:2009kt, Visinelli:2017imh, Ramberg:2019dgi}.

\item Decay of domain walls

When the primordial plasma undergoes the QCD phase transition, the effective axion potential in Eq.~\eqref{Vqcd} takes place, showing $N$ minima separated by domain walls attached to strings. Similarly to what discussed for axions from strings, there has been some controversy regarding the spectrum of axion radiated from domain walls. Ref.~\cite{Nagasawa:1994qu} claims that the energy spectrum peaks around the axion mass, while in Refs.~\cite{Chang:1998tb} a larger axion population is obtained by using an emission spectrum proportional to the axion wave number. The evolution of the string-wall network with $N = 1$ has been explored in Refs.~\cite{Hiramatsu:2010yn, Hiramatsu:2012gg}, where numerical simulations have been performed to settle the controversy and a spectrum peaking at a wave number of the order of the axion mass is obtained. The contribution of cold axions from wall decay is found as $\aw = \rwt/\rmt = (32\pm 16)$~\cite{Hiramatsu:2012gg, Kawasaki:2014sqa}.

\end{itemize}

\section{Vacuum realignment mechanism} \label{sec_vrm}

The axion field originates from the breaking of the PQ symmetry at a temperature of the order of $f_A/N$. Axions, which are the quanta of the axion field, are massless from the moment of production down to the temperature of QCD transition, when the mass term in Eq.~\eqref{axionmass} turns in. In this picture, the equation of motion for the angular variable of the axion field at any time is
\be
	\ddot{\theta} + 3H\,\dot{\theta} - \frac{\bar{\nabla}^2}{a^2}\,\theta + m_A^2\,\sin\theta = 0,
	\label{eq_motion_vrm}
\ee
where $\bar{\nabla}$ is the Laplacian operator with respect to the co-moving coordinates $\bar{x}$. We re-scale time $t$ and scale factor $a$ so that these quantities are dimensionless, $t \to t/t_1$ and $a \to a/a_1$, and we use the definition in Eq.~\eqref{axionmass} for the axion mass at the time at which the coherent field oscillations begin, $m_1 = m_A(T_1) = 3H_1 = 3\beta/t_1$, as a function of the exponent $\beta$ that expresses the time dependence of the scale factor $a$ as $a = t^\beta$ in Eq.~	\eqref{at_relation}. Using this relation, we obtain $t_1 = 3\beta/m_1$, so that Eq.~\eqref{eq_motion_vrm} in these rescaled quantities reads
\be
	\ddot{\theta} + 3\frac{\dot{a}}{a}\,\dot{\theta} - \beta^2\,\frac{\nabla^2}{a^2}\,\theta + 9\beta^2\,\left(\frac{m_A}{m_1}\right)^2\,\sin\theta = 0\,.
	\label{eq_motion_vrm1}
\ee
In this last expression, we have written the Laplacian operator in terms of a rescaled co-moving spatial coordinate $x = H_1\,a_1\,\bar{x}$. Defining $\chi = 2-1/(2\beta)$, and setting
\begin{equation}
	\theta = \frac{\psi}{a^\chi},
\end{equation}
Eq.~\eqref{eq_motion_vrm1} is rewritten as
\be
	\psi'' \!+\! \chi(1-\chi)\,\frac{\psi}{a^2} \!-\! a^{4(1-\chi)} \nabla^2\,\psi + 9\left(\frac{m_A}{m_1}\right)^2 a^{\frac{3}{2\beta}}\sin\left(\frac{\psi}{a^\chi}\right) \!=\! 0,
	\label{eq_motion_vrm2}
\ee
where a prime indicates a derivation with respect to $a$. The expression above is the generalization of the equation of motion for the axion field in any cosmological model, and reduces to the usual expression in the radiation-dominated limit $\beta = 1/2$,
\be
	\psi'' - \nabla^2\,\psi + 9\left(\frac{m_A}{m_1}\right)^2\,a^3\,\sin\left(\frac{\psi}{a}\right) = 0.
	\label{eq_motion_std}
\ee
Eq.~\eqref{eq_motion_std} coincides with the results in Ref.~\cite{Kolb:1993hw}, where the conformal time $\eta$ is used as the independent variable in place of the scale factor $a$. We remark that this choice is possible in the radiation-dominated cosmology because $\eta \sim a$, whereas in a generic cosmological model this relation reads $\eta \sim a^{1/\beta-1}$ and the use of $\eta$ as the independent variable leads to a more complicated form of Eq.~\eqref{eq_motion_vrm2}. Thus, in a modified cosmology the choice of the scale factor as the independent variable leads to a simpler form of the equation of motion. Taking the Fourier transform of the axion field as
\begin{equation}
\psi(\bx) = \int e^{-i q\,\bx}\,\psi(q),
\end{equation}
we find
\be
	\psi'' + \chi(1-\chi)\,\frac{\psi}{a^2} + a^{\frac{2}{\beta}-4}\,q^2\,\psi + 9\left(\frac{m_A}{m_1}\right)^2\,a^{\frac{3}{2\beta}}\sin\left(\frac{\psi}{a^\chi}\right) = 0.
	\label{eq_motion_full}
\ee
Eq.~\eqref{eq_motion_full} expresses the equation of motion for the axion field in the variable $a$ and it is conveniently written to be solved numerically.

\subsection{Approximate solutions of the equation of motion}

Analytic solutions to Eq.~\eqref{eq_motion_full} can be obtained in the limiting regime $\theta \ll 1$, where Eq.~\eqref{eq_motion_full} reads
\be
	\psi'' + \kappa^2(a)\,\psi = 0,
	\label{eq_motion_small_phi}
\ee
with the wave number
\be
	\kappa^2(a) = \frac{\chi(1-\chi)}{a^2} + 9\left(\frac{m_A}{m_1}\,a^{\frac{1-\beta}{\beta}}\right)^2 + \left(q\,a^{\frac{1-2\beta}{\beta}}\right)^2.
	\label{definition_kappa}
\ee
An approximate solution of Eq.~\eqref{eq_motion_small_phi}, valid in the adiabatic regime in which higher derivatives are neglected, is given by setting
\be
	\psi = \psi_0(a)\,\exp\,\left(i\,\int^a\,\kappa(a')\,da'\right),
	\label{ansatz1}
\ee
where the amplitude $\psi_0$ is given by
\be
	|\psi_0(a)|^2\,\kappa(a) = {\rm const.}
\ee
Finally, an approximate solution to Eq.~\eqref{eq_motion_small_phi} is~\cite{Kolb:1993hw, Sikivie:2006ni}
\be
	\psi = \frac{\rm const.}{\sqrt{\kappa(a)}}\,\exp\,\left(i\,\int^a\,\kappa(a')\,da'\right).
	\label{ansatz2}
\ee
Each of the three terms appearing in Eq.~\eqref{definition_kappa} is the leading term in a particular regime of the evolution of the axion field. We analyse these approximate behaviour in depths in the following.

\begin{itemize}
\item Solution at early times, outside the horizon

At early times $t \sim a^{1/\beta} \lesssim t_1$ prior to the onset of axion oscillations, the mass term in Eq.~\eqref{eq_motion_small_phi} can be neglected since $m_A(a) \ll m_1$. Defining the physical wavelength $\lambda = a/q$, we distinguish two different regimes in this approximation, corresponding to the evolution of the modes outside the horizon ($\lambda \gtrsim t$) or inside the horizon ($\lambda \lesssim t$). In the first case $\lambda \gtrsim t$, Eq.~\eqref{eq_motion_small_phi} at early times reduces to
\be
	a^2\,\psi'' + \chi(1-\chi)\,\psi = 0,
	\label{motion_approx_outside}
\ee
with solution ($\theta = \psi/a^\chi$)
\be
	\theta(q,t) = \theta_1(q) + \theta_2(q)\,a^{\frac{1-3\beta}{\beta}} = \theta_1(q) + \theta_2(q)\,t^{1-3\beta}.
	\label{field_early_times}
\ee
One of the two solutions to Eq.~\eqref{motion_approx_outside} is thus a constant value $\theta_1(q)$, while the second solution drops to zero for cosmological models with $\beta > 1/3$. Regardless of the cosmological model considered, the axion field for modes larger than the horizon is ``frozen by causality''. For example, in a radiation-dominated model with $\beta = 1/2$, Eq.~\eqref{field_early_times} coincides with the result in Ref.~\cite{Sikivie:2006ni},
\be
	\theta = \theta_1(q) + \theta_2(q)\,t^{-1/2}.
	\label{field_early_times_RD}
\ee

\item Solution at early times, inside the horizon

Eq.~\eqref{eq_motion_small_phi} for modes that evolve inside the horizon $\lambda \lesssim t$ reduces to 
\be
	\psi'' + \left(q\,a^{2-1/\beta}\right)^2\,\psi = 0,
	\label{motion_approx_inside}
\ee
whose solution in a closed form, obtained through Eq.~\eqref{ansatz2} and $\theta = \psi/a^\chi$, reads
\be
	\theta = \frac{\rm const.}{a}\,\exp\,\left(iq\,\int^a\,(a')^{\frac{2\beta-1}{\beta}}\,da'\right).
	\label{solution_eqmotion_q}
\ee
The dependence of the amplitude $|\theta|\sim 1/a$ in Eq.~\eqref{solution_eqmotion_q} is crucial, since it shows that the axion number density scales with
\be
	n_A(q, t) \sim \frac{|\theta|^2}{\lambda} \sim a^{-3},
\ee
for any cosmological model considered.

\item Solution for the zero mode at the onset of oscillations

An approximate solution of Eq.~\eqref{eq_motion_small_phi} for the zero-momentum mode $q = 0$, valid after the onset of axion oscillations when $t \sim t_1$, is obtained by setting
\be
	\kappa(a) \approx 3\frac{m_A(a)}{m_1}\,a^{\frac{1-\beta}{\beta}},
\ee
so that the adiabatic solution for $\psi$ in Eq.~\eqref{ansatz2} in this slowly oscillating regime gives the axion number density
\be
	n_A^{\rm mis}(a) = \frac{1}{2}m_A(a)\,f_A^2\,\left|\frac{\psi(a)}{a^\chi}\right|^2 = \ns\,\left(\frac{a}{a_1}\right)^{-3},
	\label{numberdensity}
\ee
where $\ns$ is the number density of axions from the misalignment mechanism at temperature $T_1$,
\be
	\ns = \frac{1}{2}\,m_A(T_1)\,f_A^2\,\langle\theta_i^2\rangle.
	\label{number_density_onset}
\ee
Eq.~\eqref{numberdensity} shows that, regardless of the dominating cosmological model, the axion number density of the zero modes after the onset of axion oscillations scales with $a^{-3}$. The energy density at temperature $T_1$ is obtained as
\begin{equation}
	\rho_A^{\rm mis}(T_1) = m_A\,\ns = \frac{\Lambda_A^4\langle\theta_i^2\rangle}{2}\,\left(\frac{T_1}{\TL}\right)^{-\gamma},
	\label{energy_density_t1A}
\end{equation}
where we have used Eq.~\eqref{axionmass} to express $m_1$ in terms of $T_1$.

\end{itemize}

\section{A note on primordial black hole formation} \label{A note on primordial black hole formation} 

Primordial black holes formed through various mechanisms, of which one consists in the growing of large inhomogeneities around the QCD phase transition. The question is, should axion inhomogeneities also form black holes instead of condensing into miniclusters? To answer this question, we compute the Schwarzschild radius $r_s = M/\Mp^2$ for the primordial plasma and for the axion energy density at the onset of oscillations.

When overdensities in the primordial plasma grow larger than one, a condition for the formation of primordial black holes is met. At time $t$, the mass enclosed within a Hubble radius is $M = \rho/H^3$, and the ratio between the Schwarzschild radius $r_s$ and the horizon length $1/H$ is
\begin{equation}
	\frac{r_s}{1/H} = \frac{\rho}{H^2\,\Mp^2} =  \frac{1}{8\pi},
\end{equation}
where in the last equality we have used the Friedmann equation $H^2 = \(8\pi/3\Mp^2\)\rho$. Thus, the Schwarzschild radius is about one order of magnitude smaller than the horizon length, so a significant fraction of inhomogeneities can condense into black holes.

For axion miniclusters of radius $R_1$, the ratio is
\be
	\frac{r_s}{R_1} = \frac{H_1}{r}\,\frac{M_c}{\Mp^2} = \frac{4\pi}{3} \frac{(1+\Phi) \,r^2\,\rho_1}{H_1^2\,\Mp^2}\,,
	\label{eq:appendixC}
\ee
where in the last expression we have used Eq.~\eqref{mass_cluster} in terms of the energy density of axions $\rho_1 \equiv \at\rho_A^{\rm mis}(T_1)$. The ratio in Eq.~\eqref{eq:appendixC} is of the order of $10^{-8}$ to $10^{-13}$ for all cosmological models considered and for all physical values of $\TRH$ and $\at$. Primordial black holes cannot form from axion cold dark matter using this mechanism, mainly because the axion field is a subdominant component of the total energy density at the QCD phase transition. Recently, the production of primordial black holes from topological defects arising in the QCD axion theory has been considered in Ref.~\cite{Ferrer:2018uiu}. The constraint on scenarios in which primordial black holes accrete dark matter has been considered in Ref.~\cite{Boucenna:2017ghj}.

\bibliography{axBib}

\end{document}